\documentclass[12pt]{article}

\usepackage[english]{babel}

\usepackage[a4paper,left=2cm,right=2cm,
    top=2cm,bottom=2cm,bindingoffset=0cm]{geometry}

\usepackage{amsfonts}
\usepackage{amsmath}
\usepackage{amssymb}
\usepackage{pstricks}
\usepackage{graphicx}

\graphicspath{{./fig/}}

\newcommand{\euler}{\mathrm{e}}

\newcommand{\Roes}{R\"{o}essler}

\begin{document}

\title{Artificial neural network as a universal model of nonlinear dynamical systems}

\author{Pavel V. Kuptsov$^{1,*}$, Anna V. Kuptsova$^2$, Nataliya V. Stankevich$^1$}

\date{
  \footnotesize{
  $^1$ Laboratory of topological methods in dynamics,
  National Research University Higher School of Economics,
  Nizhny Novgorod, 25/12 Bolshay Pecherskaya str., Nizhny
  Novgorod 603155, Russia \\[2ex]
  $^2$ Institute of electronics and mechanical engineering, Yuri
  Gagarin State Technical University of Saratov, Politekhnicheskaya
  77, Saratov 410054, Russia \\[2ex]
  $^*$ Send correspondence to P.V.K. E-mail: kupav@mail.ru \\[2ex]}
  \today}

\maketitle

\begin{abstract}
  We suggest a universal map capable to recover a behavior of a wide range of
  dynamical systems given by ODEs. The map is built as an artificial neural
  network whose weights encode a modeled system. We assume that ODEs are known
  and prepare training datasets using the equations directly without computing
  numerical time series. Parameter variations are taken into account in the
  course of training so that the network model captures bifurcation scenarios of
  the modeled system. Theoretical benefit from this approach is that the
  universal model admits using common mathematical methods without needing to
  develop a unique theory for each particular dynamical equations. Form the
  practical point of view the developed method can be considered as an
  alternative numerical method for solving dynamical ODEs suitable for running
  on contemporary neural network specific hardware. We consider the Lorenz
  system, the \Roes{} system and also Hindmarch–Rose neuron. For these three
  examples the network model is created and its dynamics is compared with
  ordinary numerical solutions. High similarity is observed for visual images of
  attractors, power spectra, bifurcation diagrams and Lyapunov exponents.\\
  
\end{abstract}

\section{Introduction}

In this paper we suggest a map, i.e., a discrete time dynamical system, capable
to recover dynamics of systems given by ODEs. The map is built as an artificial
neural network whose encode the modeled system. Using neural networks for
dynamical systems reconstruction is a long standing problem. But typically
networks are used to predict dynamics when governing equations are unknown and
only time series are available~\cite{Haykin2009, Zhang2012, lewis2016deep,
  brownlee2018deep}. We assume that ODEs are known and create their neural
network model. The structure of the network is the same for all cases while
network weights are trained to fit the modeled dynamical system. To prepare
training datasets we do not use system time series. Instead we feed the network
and the modeled system by random time series sampled form normal distributions
and update network weights comparing the outputs. Parameter variations are taken
into account so that the network captures bifurcation scenarios of the modeled
system.

The motivations of this study are the following. We want to develop a method of
training an artificial neural network that can operates as a discrete time
system and can reproduce behavior of a wide variety of dynamical systems. We
consider a perceptron with one hidden layer and sigmoidal activation. Also such
network is said to consists of two dense layers. Many contemporary
investigations deal with deep networks whose number of layers is much more then
two and whose neuron interconnections is much more complicated. We prefer a
classical architecture because of a solid mathematical background behind it,
that is the universal approximation theorem. According to this theorem,
considered network is the simplest universal approximator, i.e., is able to
reproduce any function of multiple variables on a compact set. Such simple
universal model can be interesting for theoretical studies. Theoretical analysis
of a dynamical system often requires developing highly specialized mathematical
approaches unique for the system. Considering the universal model that covers a
wide range of systems one can extend the theoretical results for these range of
systems without needing always recreate a special mathematical approaches.

From the practical point of view the development of methods of creation of the
universal neural network that can models dynamics can be considered an
alternative to the existing numerical methods for solving dynamical ODEs.
Although a large variety of well established and effective numerical methods is
available for computer simulation of dynamics, these methods are basically
developed for single-thread computation. But contemporary trend in computational
hardware development is in increasing of a number of computation cores instead
of increasing of single-core speed. In particular many hardware are known today
specialized for implementing artificial neural networks. In this situation it
seems to be very important to develop new numerical approaches good fitted to a
powerful contemporary hardware. Our model operates as a neural network that can
be run either using various available today network software like
TensorFlow~\cite{tensorflow2015-whitepaper} and PyTorch~\cite{PyTorch} or it can
be downloaded to a dedicated computer chip called AI accelerator (AI stands for
artificial intelligence)~\cite{RevAlgHard2020,AiHardLitRev2021}.

\section{Mathematical background: the universal approximation theorem}

The problem of a universal construct for approximation of functions with many
variables has a long story. First can be mentioned the Weierstrass
theorem~\cite{kline1990mathematical} that states that any continuous function
over a closed interval on the real axis can be expressed in that interval as an
absolutely and uniformly convergent series of polynomials. David Hilbert in the
International Congress of Mathematicians in Paris in the year 1900 outlined 23
major mathematical problems for in the coming new century. His 13th problem is
whether solutions to 7th degree polynomial equation can be written as the
composition of finitely many two-variable functions.  Hilbert believed they
could not be. In 1956-57 years, Kolmogorov and Arnold proved that each
continuous function of $N$ variables --- including the case in which $N = 7$ ---
can be written as a composition of continuous functions of two
variables~\cite{Kolmogorov56,Kolmogorov57,Arnold57}. This is called
Kolmogorov–Arnold representation theorem.

Research interest in the virtues of multilayer perceptrons as devices for the
representation of arbitrary continuous functions was perhaps first put into
focus by Hecht-Nielsen~\cite{HechtNielson87}. In the context of traditional
multilayer perceptrons, it was Cybenko who demonstrated rigorously for the first
time that a single hidden layer is sufficient to uniformly approximate any con-
tinuous function with support in a unit hypercube~\cite{Cybenko1989}. In 1989,
two other papers were published independently on multilayer perceptrons as
universal approximators~\cite{Funahashi89, Hornik1989}. For subsequent
contributions to the approximation problem, see~\cite{Light92}. Review on this
topic can also be found in~\cite{Haykin2009}

To sum up, universal approximation theorem states that a feed-forward network
with a single hidden layer containing a finite number of neurons can approximate
continuous functions on compact subsets of $\mathbb{R}^N$, under mild
assumptions on the activation function. The theorem thus states that simple
neural networks can represent a wide variety of interesting functions when given
appropriate parameters.

\section{The network and training details}

Assume that we have ODE
\begin{equation}
  \label{eq:ode_common}
  \dot u = f(u, p),
\end{equation}
where $u\in \mathbb{R}^{N_u}$ is a vector of $N_u$ dynamical variables, and
$p\in \mathbb{R}^{N_p}$ is a vector of $N_p$ parameters.

We consider a perceptron with one hidden layer, or using more contemporary
terms, a network with two dense layers. Formally the network can be represented
as a function that maps vectors $u\in \mathbb{R}^{N_u}$ to vectors
$d\in \mathbb{R}^{N_u}$,
\begin{equation}
  \label{eq:network_formal}
  d = F(u, p, w),
\end{equation}
where $w$ is a vector of network weights. Our purpose is to tune $w$ in such a
way that
\begin{equation}
  \label{eq:netw_idea1}
  u(t+\Delta t)=u(t) + d(t)
\end{equation}
where $u(t)$ is a solution to ODE~\eqref{eq:ode_common} and $\Delta t$ is a time
step. The size of the time step is defined before training the network. We take
$\Delta t=0.01$.

Consider a semi implicit numerical scheme of ODE solution:
\begin{equation}
  \label{eq:netw_idea2}  
  u(t+\Delta t) = u(t) + \frac{\Delta t}{2} \{ f[u(t)] + f[u(t+\Delta t)] \}
\end{equation}
Compute the difference between~\eqref{eq:netw_idea1} and \eqref{eq:netw_idea2}:
\begin{equation}
  \label{eq:netw_idea3}
  e = d(t) - \frac{\Delta t}{2} \{ f[u(t)] + f[u(t+\Delta t)] \},
\end{equation}
where $e$ is the approximation error. Substituting $u(t+\Delta t)$ as
$u(t)+d(t)$ from Eq.~\eqref{eq:netw_idea1} and omitting $t$ we obtain
\begin{equation}
  \label{eq:netw_idea4}
  e = d - \frac{\Delta t}{2} [f(u) + f(u + d)]
\end{equation}
The network approximation~\eqref{eq:network_formal}, \eqref{eq:netw_idea1} works
well if the approximation error tends to zero $e\to 0$ for any $u$ and $p$ from
the domain of interest.

Before training we need to define a localization areas for $u$ and $p$. This is
done empirically via testing various numerical solutions of
Eq.~\eqref{eq:ode_common}. We define in this way a mean value $\mu_u$ and a
standard deviation $s_u$ of $u$ and the corresponding $\mu_p$ and $s_p$ varying
parameters $p$. The training occurs on a random $u$ and $p$ sampled from normal
distribution defined by given mean values $\mu_u$, $\mu_p$ and standard
deviations $s_u$ and $s_p$. Since we use a random number generator to produce
dataset its size is limited only by a period of random number generator that is
very large.

Let us now discuss the structure of the network denoted above as $F(u,p,w)$.
The network includes liner and nonlinear data transformations. The linear one is
done via multiplication of data vectors by a matrix of neuron weights. For
neural networks the usual order of vector-matrix manipulation is the reversed:
Typically we multiply a matrix by a vector-column and in the neural network
context a vector-row is multiplied by a matrix. This is done because in the
course of training a batch of vectors is processed in parallel. A rectangular
data matrix with the vectors stowed in rows is multiplied by a matrix of
weights. Thus we assume that $u$ and $p$ are vector-rows of dimension $N_u$ and
$N_p$ respectively.

The training data vectors $u$ and $p$ are sampled from a normal distribution and
elements of $u$ and $p$ can have different scales. Thus the first transformation
of the network inputs $u$ and $p$ is a non-trainable normalization layer that
rescales inputs to a standard normal distribution.
\newcommand{\layNorm}{\mathop{\mathrm{Norm}}}
\newcommand{\layDenorm}{\mathop{\mathrm{Denorm}}}
\begin{gather}
  \label{eq:norm_layer}
  \layNorm(x) = (x - \mu_x) / s_x, \\
  \label{eq:denorm_layer}
  \layDenorm(x) = x s_x + \mu_x.
\end{gather}
Here $x$, $\mu_x$ and $s_x$ are vectors-rows and operations are performed
element-wise. Also we define here the layer performing backward operation
$\layDenorm()$. It will be done at the very end of the network to fit the values
to an appropriate range. It might be seem that the layers~\eqref{eq:norm_layer}
and \eqref{eq:denorm_layer} are superfluous --- one can expect that the network
is able to fit these scales itself in the course of training. But in fact this
is not the case. All network training methods are developed in the assumption
that both inputs and outputs do not deviate much from a standard range. So the
training is efficient if we know in advance what are the ranges of the inputs
and the outputs and rescale them appropriately.

After normalization we concatenate two resulting row vector into the one vector:
\newcommand{\layConcat}{\mathop{\mathrm{Concat}}}
\begin{equation}
  \label{eq:concat_layer}
  \layConcat(x, y) = (x, y)
\end{equation}
Here $x$ and $y$ are vectors of $N_x$ and $N_y$ elements, respectively, and
$(x,y$) is a row vector of $N_x+N_y$ elements.

The next step is a dense layer. This is mere a affine transformation:
\newcommand{\layDense}{\mathop{\mathrm{Dense}}}
\begin{equation}
  \label{eq:dense_layer}
  \layDense(x, N) = x W_{x,N} + b_{x,N}
\end{equation}
Here $W_{x,N}$ is a rectangular matrix whose number of rows equals to the number
of columns of $x$ and the number of columns of $W_{x,N}$ is $N$, $b_{x,N}$ is a
vector-row with $N$ elements.

After that a nonlinear transformation is applied that is called activation:
\newcommand{\layActiv}{\mathop{\mathrm{Activ}}}
\begin{equation}
  \label{eq:activ_layer}
  \layActiv(x) = \sigma(x)
\end{equation}
Here $\sigma()$ is a scalar function of a scalar argument and if a vector is passed
to it the element-wise operation is assumed.

Subsequent transformations are done using already defined operators so that the
whole network $d=F(u,p,w)$ can be described as follows:
\begin{gather}
  \label{eq:network1}
  z = \layConcat(\layNorm(u), \layNorm(p)) \\
  \label{eq:network2}
  h = \layActiv(\layDense(z, N_h)) \\
  \label{eq:network3}
  g = \layDense(h, N_u) \\
  \label{eq:network4}
  d = \Delta t \layDenorm(g)
\end{gather}
Variable $w$ in $F(u,p,w)$ represents a set of trainable parameters of the
networks. As follows from the equations above
\begin{equation}
  \label{eq:network_param}
  w = \left\{
    W_{z,N_h}, b_{z,N_h}, W_{h,N_u}, b_{h,N_u}
  \right\}  
\end{equation}

In the very beginning the network parameters $w$ are initialized at random. Then
the training process is performed as follows.  We generate an input batch
$\{U, P\}$ of $N_{\text{batch}}$ random $u$ and $p$ sampled form a normal
distribution. Here $U$ and $P$ matrices with $N_{\text{batch}}$ rows and their
number of columns are $N_u$ and $N_p$, respectively. This batch is feed to the
network \eqref{eq:network1}-\eqref{eq:network4} and the matrix $D$ with
$N_{\text{batch}}$ rows and $N_u$ columns is obtained. Then the input matrix $U$
and the output one $D$ is substituted into \eqref{eq:netw_idea4} to compute an
error matrix $E$ of $N_{\text{batch}}$ rows and $N_u$ columns. Finally a mean
squared error (MSE) is computed for the elements of $E$ as:
\begin{equation}
  \label{eq:network_loss}
  \ell = \frac{1}{N_{\text{batch}} N_u} \sum_{i=1}^{N_{\text{batch}}}\sum_{j=1}^{N_u} e_{ij}^2
\end{equation}
This $\ell$ is the loss function for our training. To update the network
parameters a gradient of $\ell$ is computed with respect of each of the network
parameter gathered in $w$, see \eqref{eq:network_param}, and then it is used in
a gradient descent step that computes corrections to the network parameters with
respect to the minimization of $\ell$. The simplest version of the gradient
descent step reads
\begin{equation}
  \label{eq:grad_descent}
  w \leftarrow w - \gamma \nabla_w \ell
\end{equation}
where the step size scale $\gamma$ is a small parameter controlling the
convergence.

The iteration that starts from a random batch generation and ends after updating
the network parameters via the gradient descent is repeated $t_{\text{epoch}}$
times. This is considered as an epoch. Notice that usually the epoch has a bit
different meaning. A neural network is trained on a large unaltered dataset it
cannot not be passed to the network at once due to the lack of a computer
memory. In this case the whole dataset is split into batches (they are also
called mini-batches) and they are passed one by one. The parameter updates are
computed for each batch. The optimization method applied not to the whole
dataset at once but to its batches is called stochastic gradient descent and the
epoch ends when all the bathes have shown to the network. In our case the
batches are always generated at random so that dividing training process into
epochs is required only to interrupt the training and to compute metrics to see
the progress of the network performance.

We use two metrics: the loss function~\eqref{eq:network_loss} and the mean
relative norm error (MRNE) that is defined as follows:
\begin{equation}
  \label{eq:network_mnre}
  m=\frac{1}{N_{\text{batch}}} \sum_{i=1}^{N_{\text{batch}}}
  \left(\sum_{j=1}^{N_u} |e_{ij}| / \sum_{j=1}^{N_u} |u_{ij}| \right),
\end{equation}
where $e_{ij}$ as above are the elements of the error matrix $E$ and $u_{ij}$
are the elements of the network input batch $U$. To estimate the network
performance, after each epoch we perform $t_{\text{valid}}$ validation steps:
Generate a new random batch $\{U, P\}$, feed the
network~\eqref{eq:network1}-\eqref{eq:network4}, obtain $E$ and compute
$\ell$~\eqref{eq:network_loss} and $m$~\eqref{eq:network_mnre} without network
parameters updating; finally the computed metrics are averaged over the
validation steps $t_{\text{valid}}$. The dependence of the average metrics on
the number of epochs passed is called learning curves.

For actual computations instead of the simplest one~\eqref{eq:grad_descent} we
the use more sophisticated version of the gradient descent method called
Adam. The difference is that the step size scale $\gamma$ is not a constant, but
is tuned according to the accumulated gradients on the previous
steps~\cite{kingma2014adam}. This method has a meta-parameter learning rate
$\alpha$ that control the overall scale of the computed step size. We decrease
it in the course of the computations according to the inverse time decay rule:
\begin{equation}
  \label{eq:invtimedec}
  \alpha = \frac{0.1}{1+0.96 t / (30 t_{\text{epoch}})}
\end{equation}
where $t$ is the gradient descent step, and $t_{\text{epoch}}$ is a number of
steps comprising one epoch. The particular numerical values of the coefficients
in this formula are chosen empirically to provide the fastest convergence.

At the activation layer $\layActiv()$ in Eq.~\eqref{eq:network2} we apply the
sigmoid function
\begin{equation}
  \label{eq:sigmoid}
  \sigma(x) = \frac{1}{1+\euler^{-x}}
\end{equation}

We will train the neural network models to achieve the mean relative error MRNE
at level $10^{-5}$.

The transformation that is done by the network under
consideration~\eqref{eq:network_formal}, \eqref{eq:network1}-\eqref{eq:network4}
can be represented as a map. Normalization operator in Eq.~\eqref{eq:network1}
can be taken into account inside the dense layer in \eqref{eq:network2} by an
appropriate rescaling and shift of the elements of $W_{z,N_h}$ and $b_{x,N_h}$.
Similarly, denormalization operator in Eq.~\eqref{eq:network4} can be merged
with the dense layer in \eqref{eq:network3}. Also instead of concatenating the
normalized vectors $u$ and $p$ we split the matrix $W_{z,N_h}$ into two blocks
corresponding to $u$ and $p$ respectively. As a result we obtain the following
map that models solutions to Eq.~\eqref{eq:ode_common}:
\begin{equation}
  \label{eq:ode_netw_sol}
  u_{n+1} = u_n + \sigma(u_n A_0 + p B_0 + a_0) A_1 + a_1
\end{equation}
where $A_0$ is a matrix with $N_u$ rows and $N_h$ columns, $B_0$ has $N_p$ rows
and $N_h$ columns, $A_1$ is a matrix with $N_h$ rows and $N_u$
columns. Vector-row $a_0$ has $N_h$ elements and $a_1$ has $N_u$ elements.

Equation~\eqref{eq:ode_netw_sol} is a universal model of a solution to
ODE~\eqref{eq:ode_common}. Particular system is selected by choosing an
appropriate size $N_h$ of the hidden layer and by numerical values of the
elements of matrices $A_0$, $B_0$, $A_1$ and vectors $a_0$ and $a_1$.

For Eq.~\eqref{eq:ode_netw_sol} we can find the variational equation suitable
for applying to this system the Lyapunov analysis, in particular, for computing
Lyapunov exponents. Differentiating the elements of $u_{n+1}$ by the elements
$u_n$ one obtains the Jacobian matrix:
\begin{equation}
  \label{eq:ode_netw_jac}
  J_n = I + A_0 H_n A_1
\end{equation}
where $I$ is the identity matrix, and $H_n$ is a diagonal square matrix $N_h$ by
$N_h$:
\begin{equation}
  \label{eq:diag_matr_h}
  H_n = \mathop{\mathrm{diag}}(h_n (1-h_n))
\end{equation}
and $h_n$ is a row-vector computed as
\begin{equation}
  h_n = \sigma(u_n A_0 + p B_0 + a_0)  
\end{equation}
In the other words it is computted according to Eqs.~\eqref{eq:network1},
\eqref{eq:network2} when $u_n$ and $p$ corresponding to the current trajectory
point are substituted there.

Thus the variational equation for the system~\eqref{eq:ode_netw_sol} reads:
\begin{equation}
  \label{eq:ode_netw_var}
  \delta u_{n+1} = (I + A_0 H_n A_1) \delta u_n
\end{equation}

This variational equation can be used to compute Lyapunov exponents. For this
purpose we apply the standard algorithm~\cite{Benettin80,Shimada79}: Iterate the
main system~\eqref{eq:ode_netw_sol} simultaneously with the required number of
copies of the variational Eq.~\eqref{eq:ode_netw_var} with periodic
orthogonalization and normalization of the set of vectors $\delta
u_n$. Accumulated and averaged in time logarithms of the norms of variational
vectors converge to the Lyapunov exponents.

Since the training and running of neural network is usually done in a
multithread computation environment the preferable way of finding of the
exponents is to iterated vary many trajectories simultaneously for not very
large time cuts and then average the resulting exponents over the trajectories.

All the computations including training and running are preformed using
TensorFlow~\cite{tensorflow2015-whitepaper} and CUDA~\cite{CUDA} software.

\section{Models}

\subsection{Lorenz system}

First we consider the Lorenz system~\cite{Lorenz63,Sparrow82,SchusJust05}:
\begin{equation}
  \label{eq:lorenz_ode}
  \begin{gathered}
    \dot x = \sigma(y-x),\\
    \dot y = x(r-z)  - y,\\
    \dot z = xy - bz
  \end{gathered}
\end{equation}
To train this model we choose the vectors of mean $\mu_u$
and standard deviation $s_u$ as follows:
\begin{equation}
  \label{eq:lorenz_scl}
  \begin{gathered}
    \mu_u = (0, 0, 0), \; s_u = (10, 10, 20) \\
    \mu_p = (0, 0, 0), \; s_p = (5, 20, 2) \\
    \mu_g = (0, 0, 0), \; s_g = (70, 280, 110)
  \end{gathered}
\end{equation}
These vectors are used in Eq.~\eqref{eq:network1}, see also
Eq.~\eqref{eq:norm_layer}.

The vectors $\mu_g$ and $s_g$ are computed as mean and standard deviation of
elements of $f(u,p)$ that is the right hand side of Eq.~\eqref{eq:lorenz_ode}
when $u$ and $p$ are sampled from normal distribution with mean and standard
deviations $\mu_u$, $s_u$, $\mu_p$ and $s_p$. In this case the network output
$d$, see Eq.~\eqref{eq:network4} will approximately have the range of
$\frac{\Delta t}{2} [f(u) + f(u + d)]$, see Eqs.~\eqref{eq:netw_idea2}
and~\eqref{eq:netw_idea3}.

\begin{figure}
  \centering
  \includegraphics[width=0.96\columnwidth]{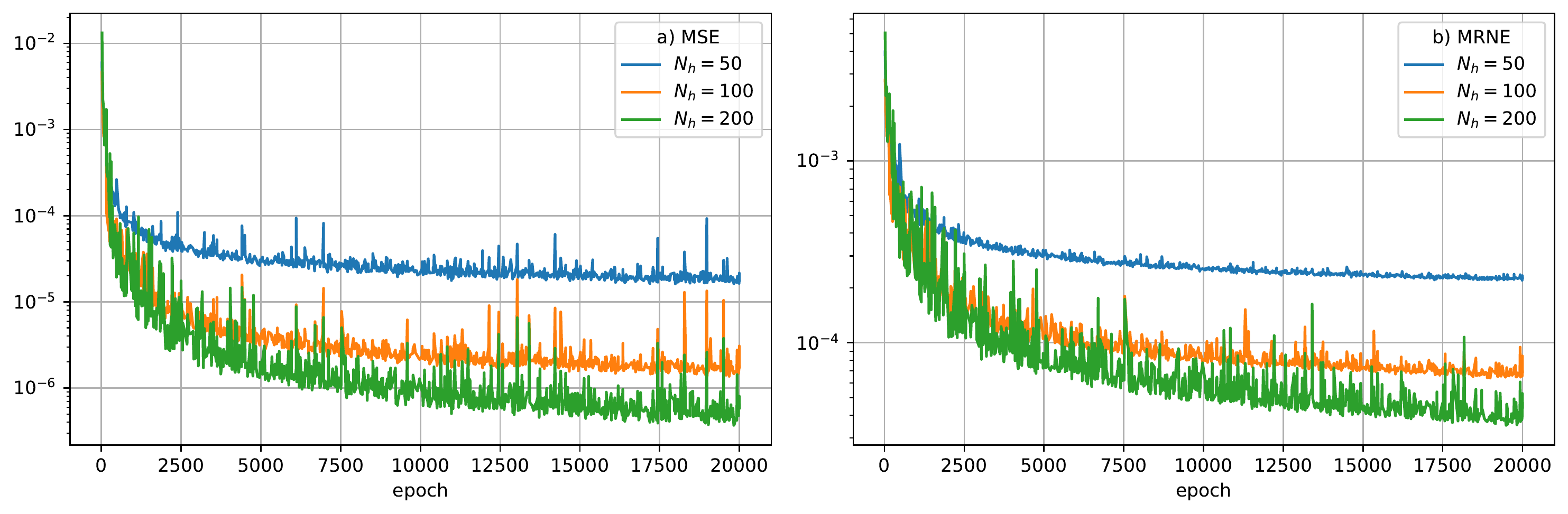}
  \caption{\label{fig:lorenz_learn}Learning curves for the Lorenz network model
    corresponding to ODEs~\eqref{eq:lorenz_ode}. The curves are computed at
    validation steps, i.e., for inputs that did not used for updating the
    network parameters. Panel (a) represents MSE,
    Eq.~\eqref{eq:network_loss}. Panel (b) shows MRNE,
    Eq.~\eqref{eq:network_mnre}. Different curves correspond to different
    $N_h$.}
\end{figure}

One more parameter that we need to define is $N_h$, the size of the hidden
layer, see~\eqref{eq:network2}. We consider different values to check which one
is preferred. Figure~\ref{fig:lorenz_learn} shows the learning curves for the
Lorenz system. Its panels (a) and (b) show MSE~\eqref{eq:network_loss} and
MRNE~\eqref{eq:network_mnre}, respectively, computed for validation data. Three
cases are shown corresponding to $N_h=50$, $100$ and $200$. We see that all
there curves decay that means that the performance of the network improves. The
fastest decay is observed for $N_h=200$. In what follows we will consider the
network with $N_h=200$. In the course of the training after each epoch we
compare the attained MRNE level with the previously smallest. And if the new one
is smaller, we save the corresponding network parameters $w$, see
Eq.~\eqref{eq:network_param}. Up to 20000 epoch we were able to find a network
whose MRNE is approximately $3\times 10^{-5}$. We use this metric as a criterion
of the performance because it is normalized by the dynamical variables scale so
that we can compare the performance of different systems.  Since MRNE is already
sufficiently small we did not considered larger values of $N_h$.

\begin{figure}
  \centering
  a)\includegraphics[width=0.46\columnwidth]{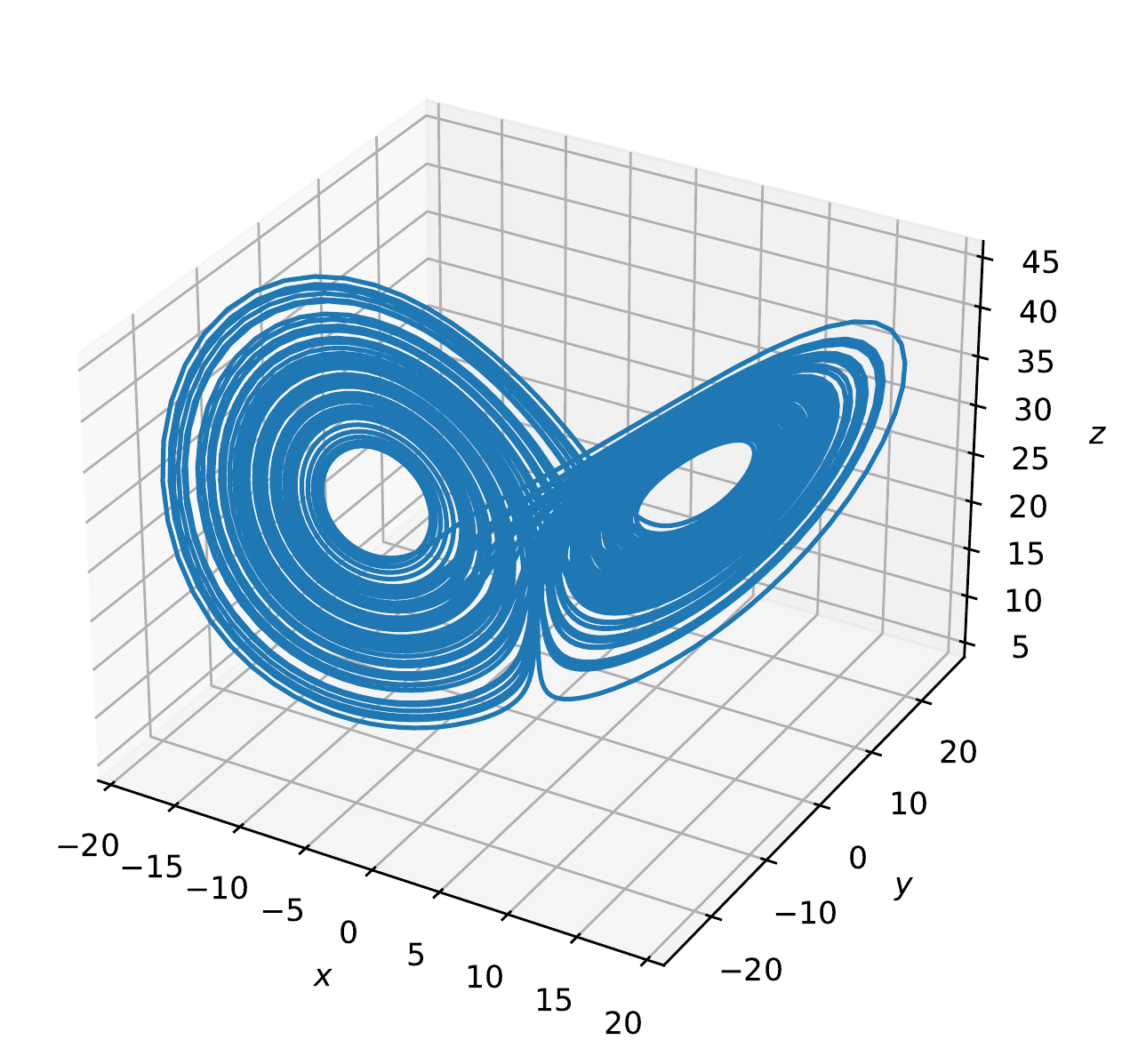}
  b)\includegraphics[width=0.46\columnwidth]{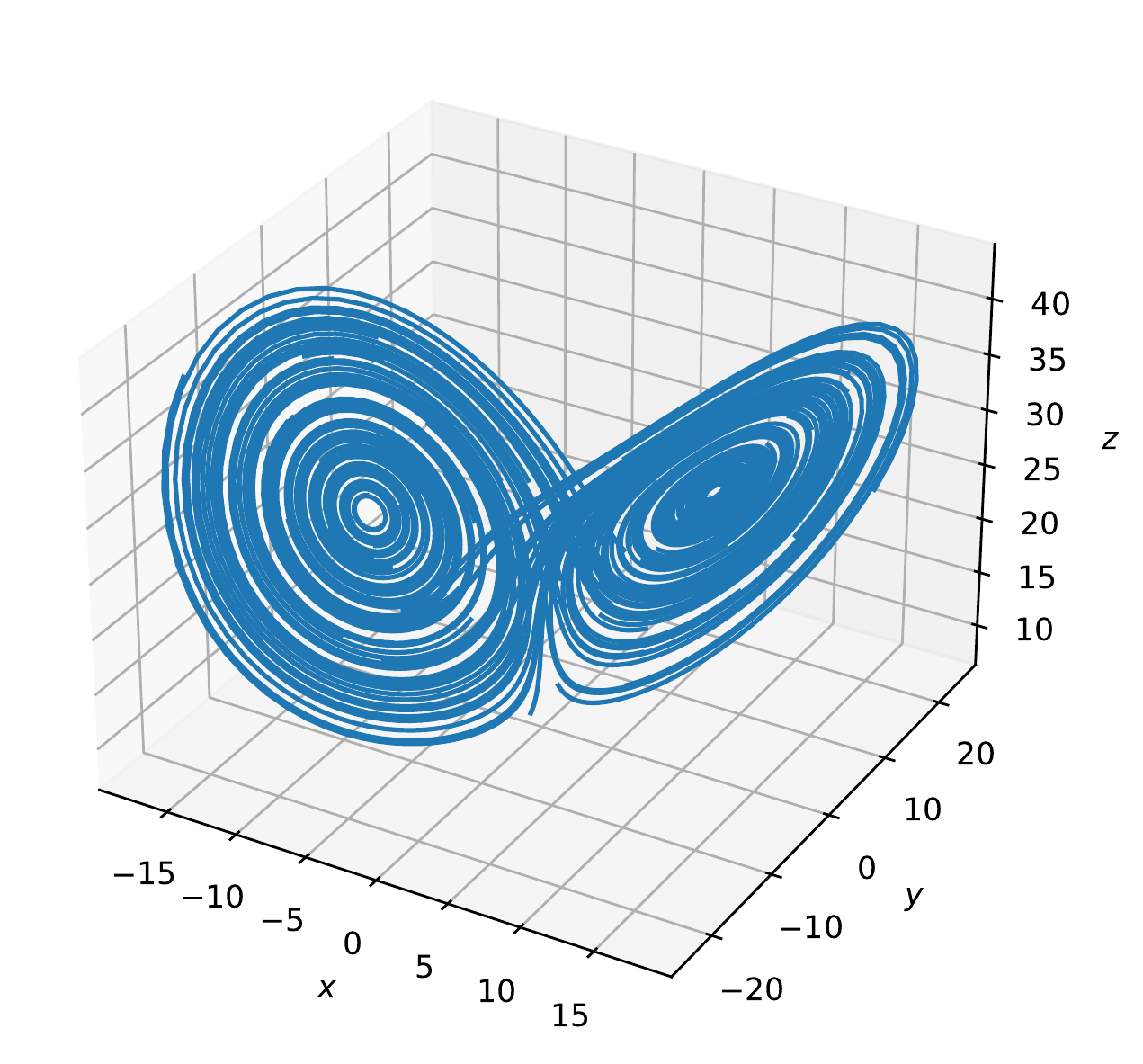}
  \caption{\label{fig:lorenz_attr}Lorenz attractor computed (a) as a numerical
    solution of Eqs.~\eqref{eq:lorenz_ode} using the forth order Runge-Kutta
    method and (b) as iteration of the network
    model~\eqref{eq:ode_netw_sol}. Parameters are $\sigma=10$, $r=28$, $b=8/3$.}
\end{figure}

\begin{figure}
  \centering
  a)\includegraphics[width=0.46\columnwidth]{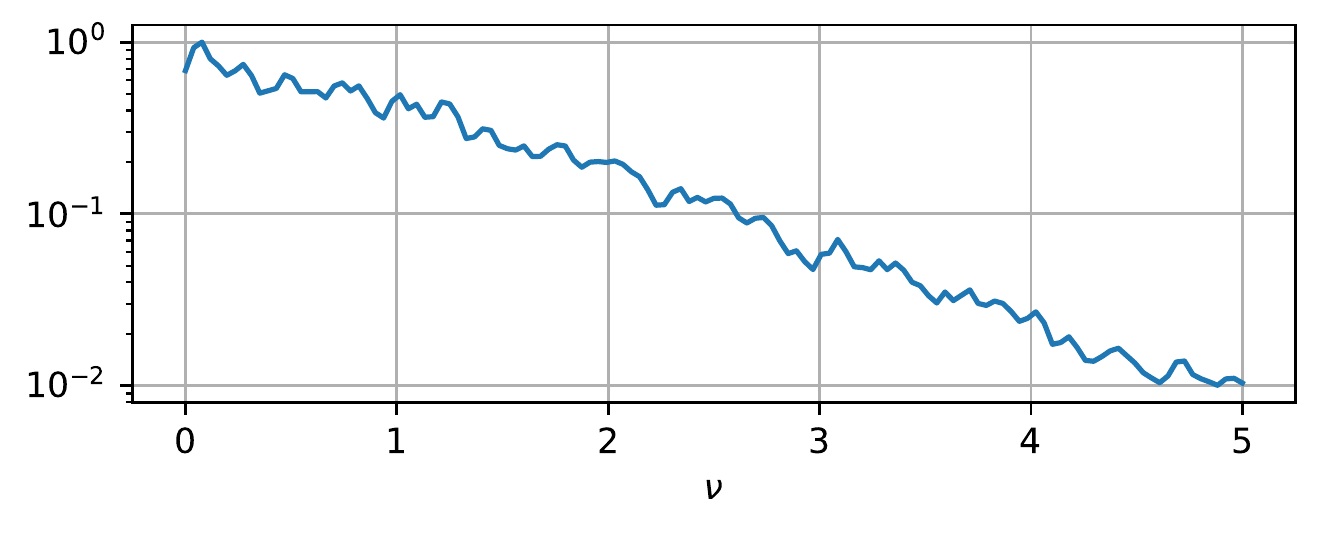}
  b)\includegraphics[width=0.46\columnwidth]{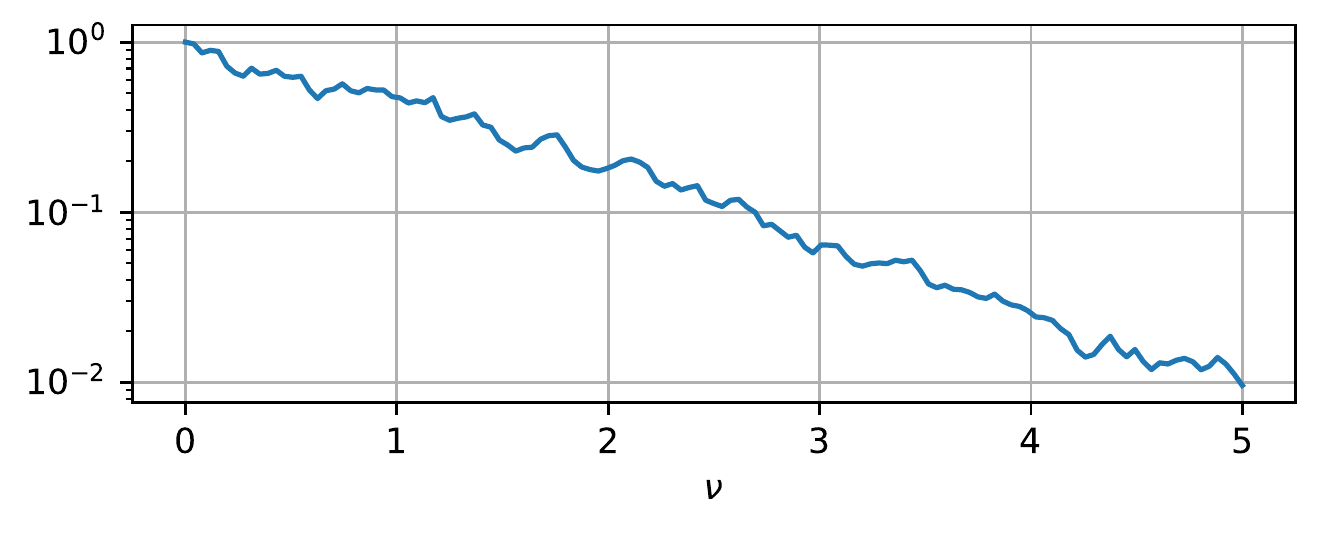}\\
  c)\includegraphics[width=0.46\columnwidth]{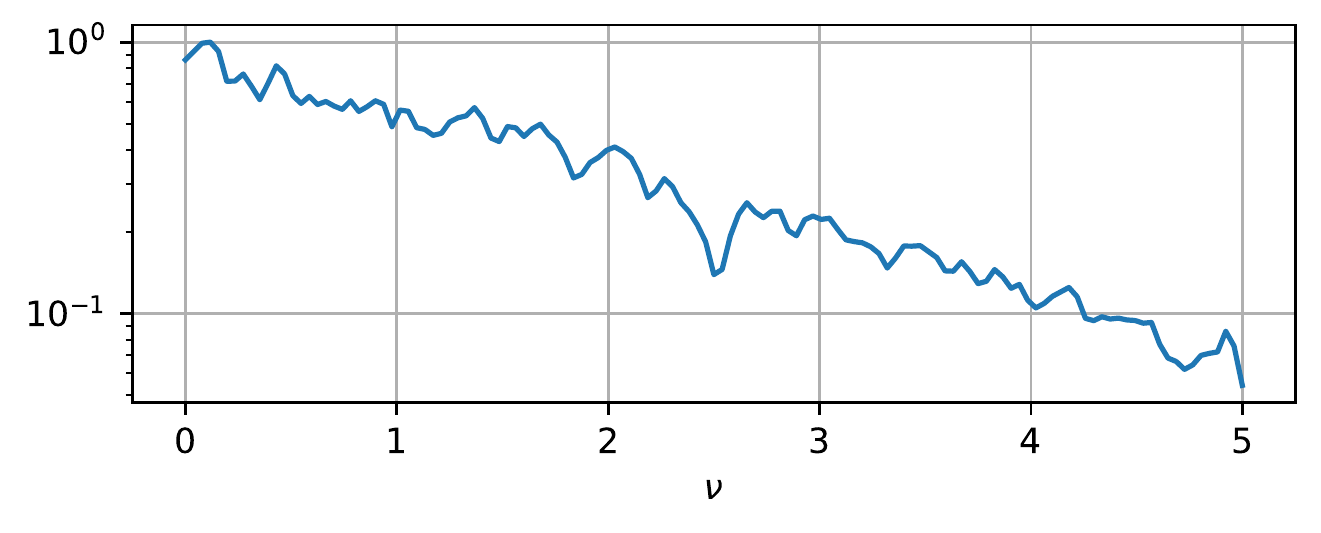}
  d)\includegraphics[width=0.46\columnwidth]{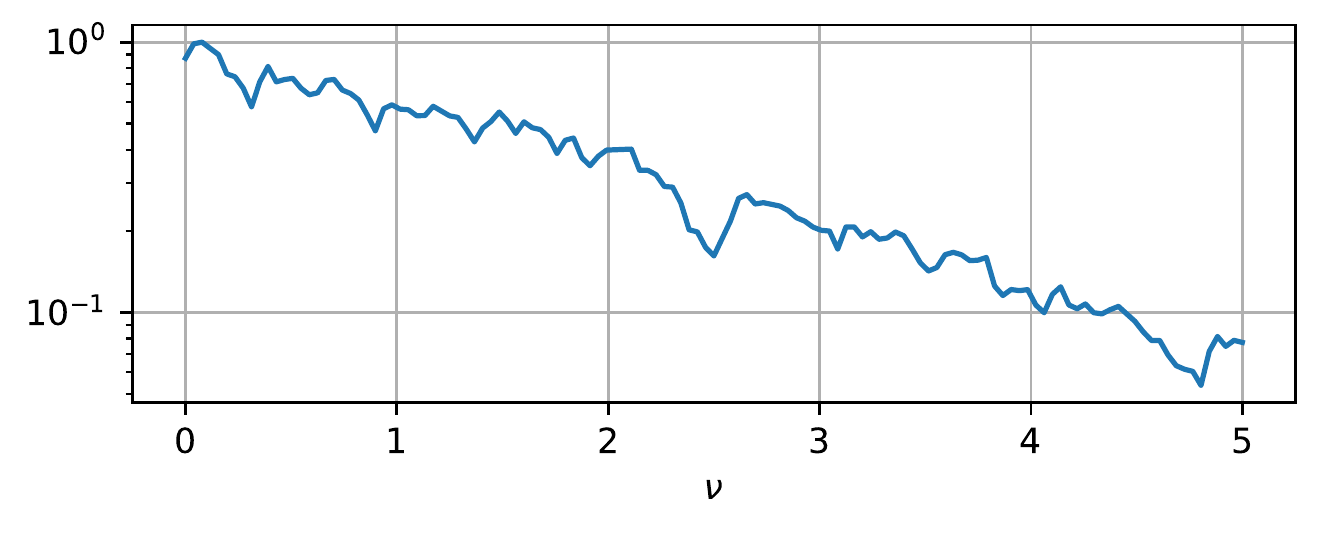}  
  \caption{\label{fig:lorenz_four}Fourier spectra of the Lorenz attractor. Data
    series for panels (a) and (c) are computed numerically using the forth order
    Runge-Kutta method, and data for the panels (b) and (d) are obtained after
    iterations of the network model~\eqref{eq:ode_netw_sol}.  Parameter for the
    panels (a) and (b) are $\sigma=10$, $r=28$, $b=8/3$, and panels (c) and (d)
    are obtained with $\sigma=16$, $r=45.92$, $b=4$}
\end{figure}

The training result is shown in Figs.~\ref{fig:lorenz_attr} and
\ref{fig:lorenz_four}. Figure \ref{fig:lorenz_attr}(a, b) demonstrates the Lorenz
attractor computed for the standard set of parameters $\sigma=10$, $r=28$,
$b=8/3$ using fourth order Runge-Kutta method (a) and the network model
\eqref{eq:ode_netw_sol} (b). Observe very high coincidence of two plots.

Neural networks architecture is very good suited for parallel computations. So
doing computations with the network model we employ it considering multiple
trajectories at once: to plot Fig.~\ref{fig:lorenz_attr}(a) via the Runge-Kutta
method we compute $10000$ steps with the time interval $\Delta t=0.01$, while in
Fig.~\ref{fig:lorenz_attr}(b) we compute $100$ trajectories at once, each of the
length $100$ steps $\Delta t=0.01$.

Figure~\ref{fig:lorenz_four} shows Fourier spectra computed for $x$ at the
parameters $\sigma=10$, $r=28$, $b=8/3$, panels (a, b), and $\sigma=16$,
$r=45.92$, $b=4$, panels (c, d). Left panels (a) and (c) are computed for the
Runge-Kutta data and the right ones are obtained for the network model. The
spectra coincide very well that indicates that the obtained
network~\eqref{eq:ode_netw_sol} models the Lorenz dynamics very well.

Now compute Lyapunov exponents using the standard
algorithm\cite{Benettin80,Shimada79}. Using the Runge-Kutta method and at
$\sigma=10$, $r=28$, $b=8/3$ we obtain the values of $\lambda_i$ in
Eq.~\eqref{eq:lorenz_lyap_ode1}. Lyapunov exponents $\tilde\lambda_i$ computed
for the network model \eqref{eq:ode_netw_sol} and corresponding variational
equation \eqref{eq:ode_netw_var} are in Eq.~\eqref{eq:lorenz_lyap_ntw1}.
Observe the very good coincidence. Notice that $\lambda_2$ is expected to be
zero since describes marginally stable perturbations along trajectories. However
actual values in computations are never exact zero. Their closeness to zero
indicates the quality of the computation. In our case both $\lambda_2$ and
$\tilde\lambda_2$ are very small.

\begin{align}
  \label{eq:lorenz_lyap_ode1}
  \lambda_1&=0.906 & \lambda_2&=8.26\times 10^{-6} & \lambda_3&=-14.6 \\
  \label{eq:lorenz_lyap_ntw1}  
  \tilde\lambda_1&=0.905 & \tilde\lambda_2&=1.26\times 10^{-5} & \tilde\lambda_3&=-14.6
\end{align}

Similarly the Lyapunov exponents are computed for the parameters $\sigma=16$,
$r=45.92$, $b=4$. Observe again the very high similarity of $\lambda_i$ with
network model exponents $\tilde\lambda_2$:
\begin{align}
  \label{eq:lorenz_lyap_ode2}
  \lambda_1&=1.50 & \lambda_2&=-1.89\times 10^{-5} & \lambda_3&=-22.5 \\
  \label{eq:lorenz_lyap_ntw2}  
  \tilde\lambda_1&=1.49 & \tilde\lambda_2&=4.34\times 10^{-5} & \tilde\lambda_3&=-22.7
\end{align}

\subsection{\Roes{} system}

Another system that we consider is the \Roes{}
system~\cite{Roess76,SchusJust05,KuzDynChaos06}:
\begin{equation}
  \label{eq:roessler_ode}
  \begin{gathered}
    \dot x = -y - z,\\
    \dot y = x + a y,\\
    \dot z = b + z (x-c),
  \end{gathered}
\end{equation}
For this system we choose the following $\mu_{u,p}$ and $s_{u,p}$ and compute
the corresponding $\mu_g$ and $s_g$:
\begin{equation}
  \label{eq:roessler_scl}
  \begin{gathered}
    \mu_u = (0,0,0), \; s_u = (10,10,10), \\
    \mu_p = (0,0,0), \; s_p = (10,10,10), \\
    \mu_g = (0,0,0), \; s_g = (14, 101, 142).
  \end{gathered}
\end{equation}

\begin{figure}
  \centering
  \includegraphics[width=0.96\columnwidth]{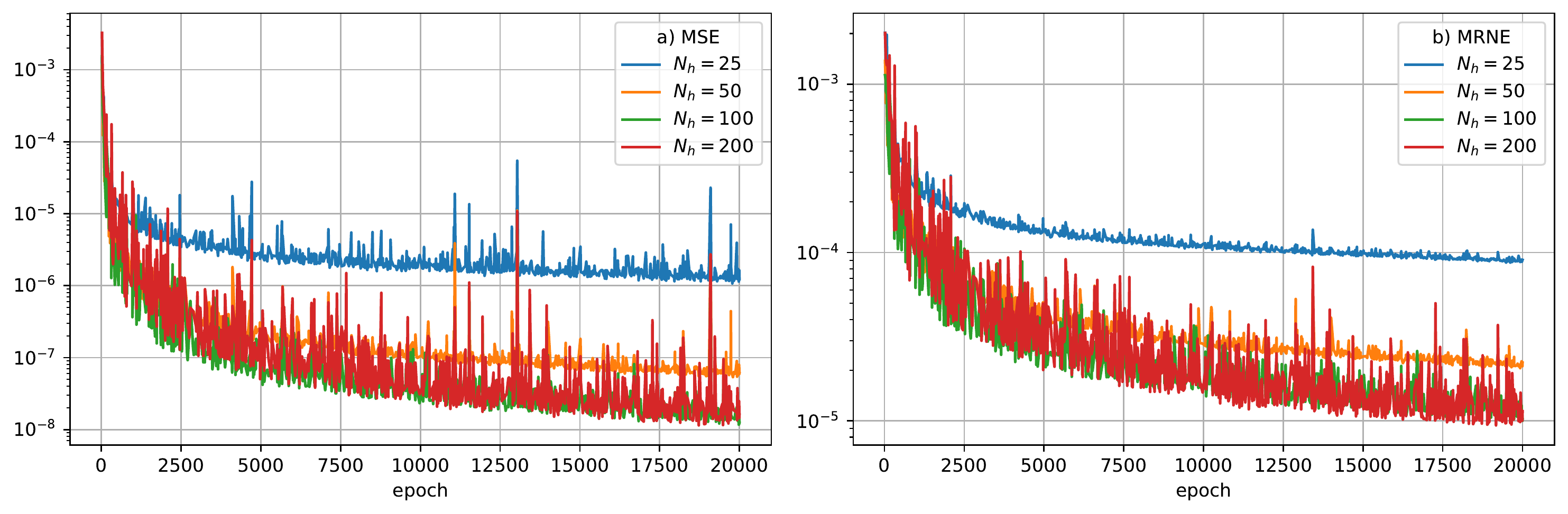}
  \caption{\label{fig:roessler_learn}Learning curves for the \Roes{} network
    model corresponding to ODEs~\eqref{eq:roessler_ode}.}
\end{figure}

Figure~\ref{fig:roessler_learn} demonstrates the learning curves for the \Roes{}
system. We observe that the training now goes much faster then for the Lorenz
system, see Fig.~\ref{fig:lorenz_learn}. Inspecting the learning curves we can
conclude that the network models with $N_h=100$ and $200$ do not differ
much. So, unlike the Lorenz system we will consider the network model for the
\Roes{} system with $N_h=100$. After 20000 epochs of the training we save the
model with the smallest MRNE equals to approximately $1.0\times 10^{-5}$.

\begin{figure}
  \tabcolsep=0pt
  \begin{center}
  \begin{tabular}{cc}
    \begin{minipage}{0.5\textwidth}
      a)\includegraphics[width=0.9\columnwidth]{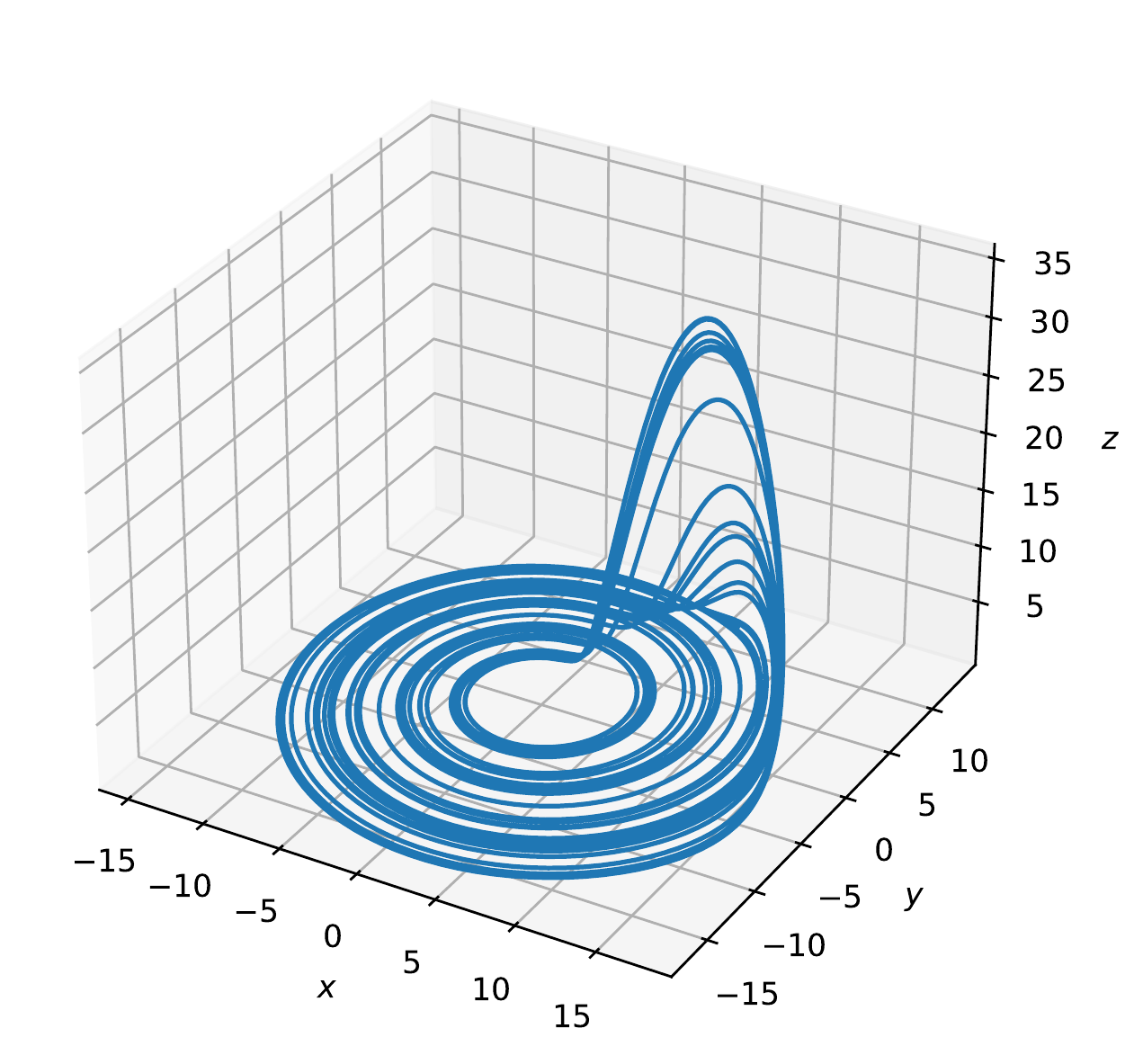}\\
      b)\includegraphics[width=0.9\columnwidth]{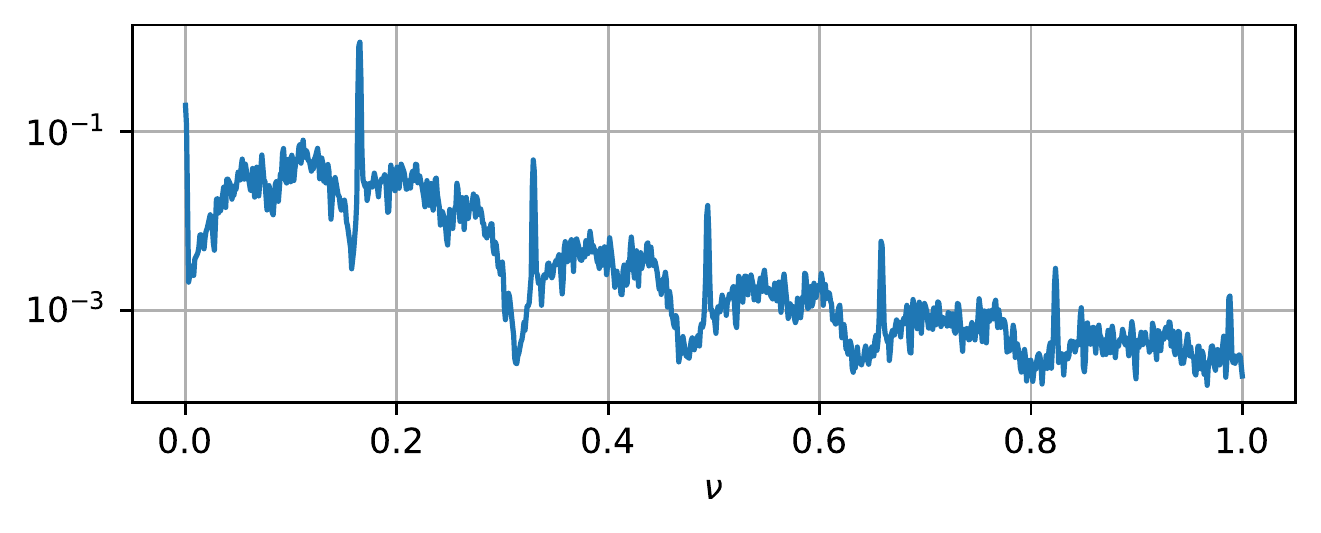}
    \end{minipage} &
    \begin{minipage}{0.5\textwidth}
      c)\includegraphics[width=0.9\columnwidth]{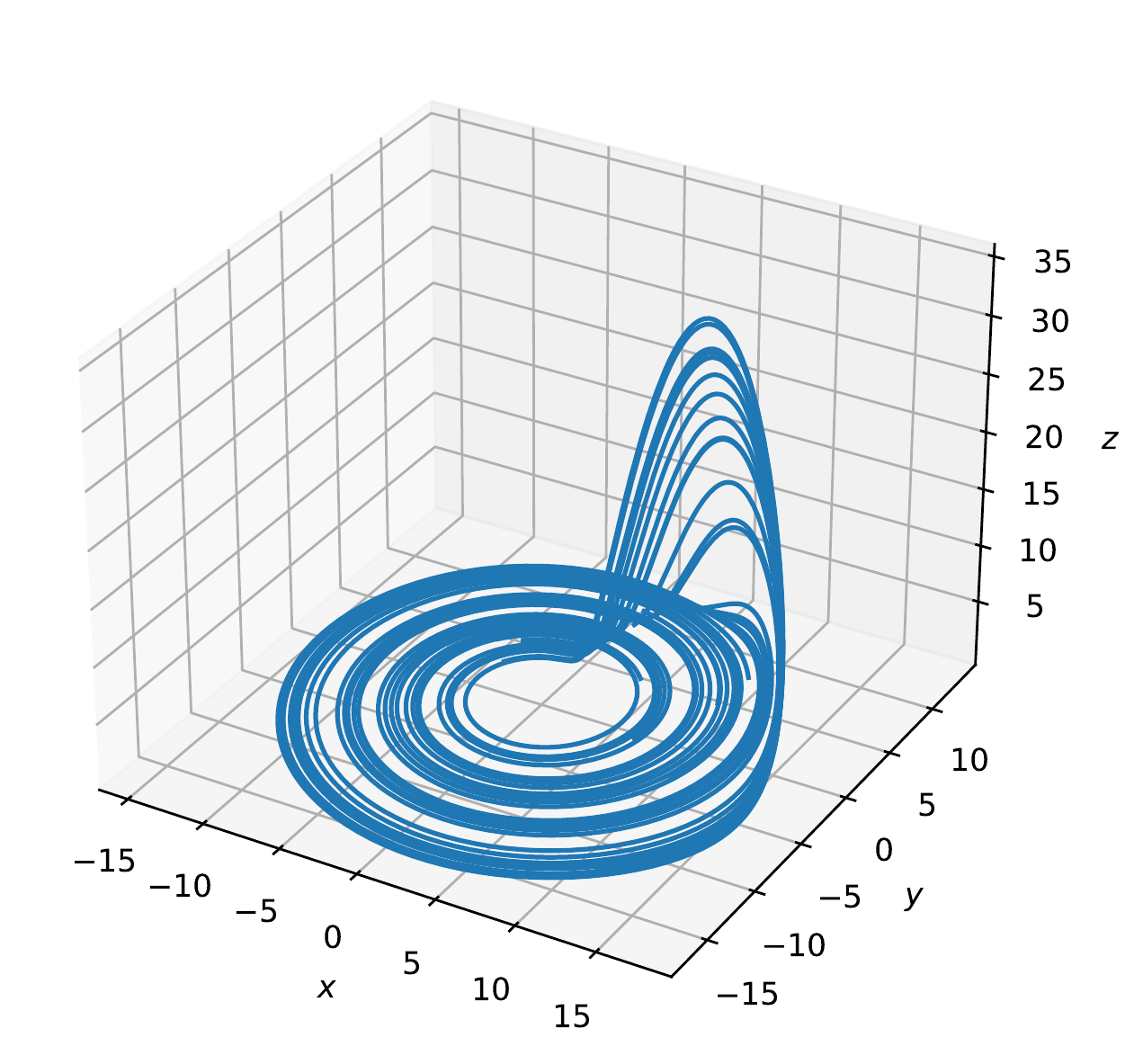}\\
      d)\includegraphics[width=0.9\columnwidth]{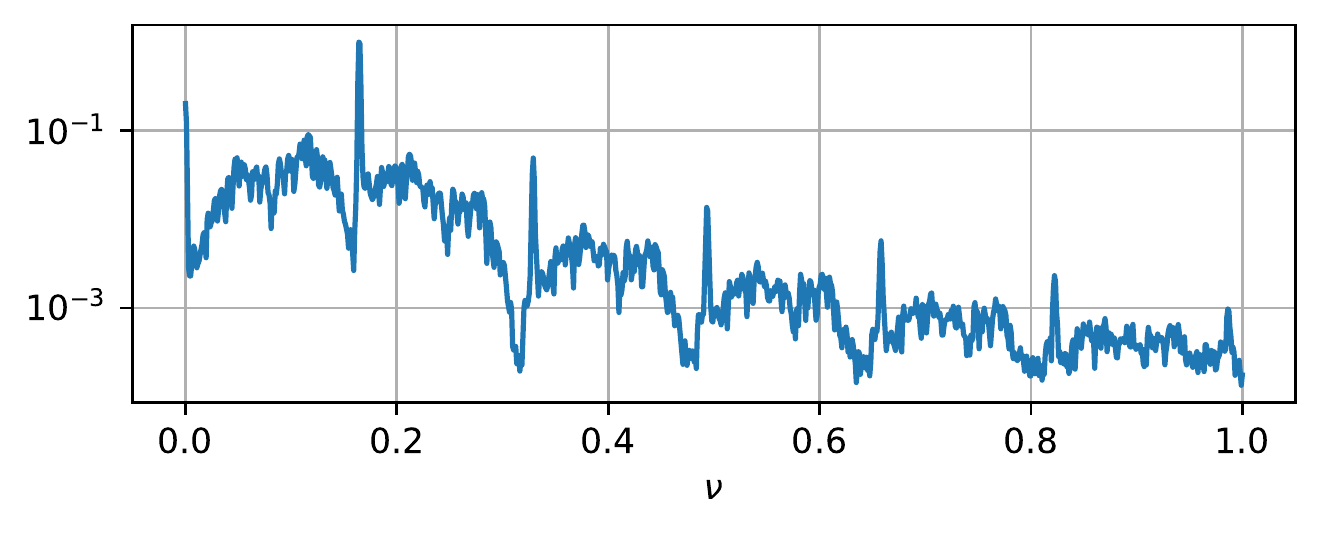}
    \end{minipage}
  \end{tabular}
  \end{center}  
  \caption{\label{fig:roessler_attr_four_chaos}\Roes{} chaotic attractor at
    $a=0.15$, $b=0.2$, and $c=10$, panels (a) and (c) and the corresponding
    Fourier spectra, panels (b) and (d). Panels (a) and (b): numerical solution
    of ODEs~\eqref{eq:roessler_ode}. Panels (c) and (d): iterations of a trained
    network model.}
\end{figure}

\begin{figure}
  \begin{center}
  \begin{tabular}{cc}
    \begin{minipage}{0.5\textwidth}
      a)\includegraphics[width=0.9\columnwidth]{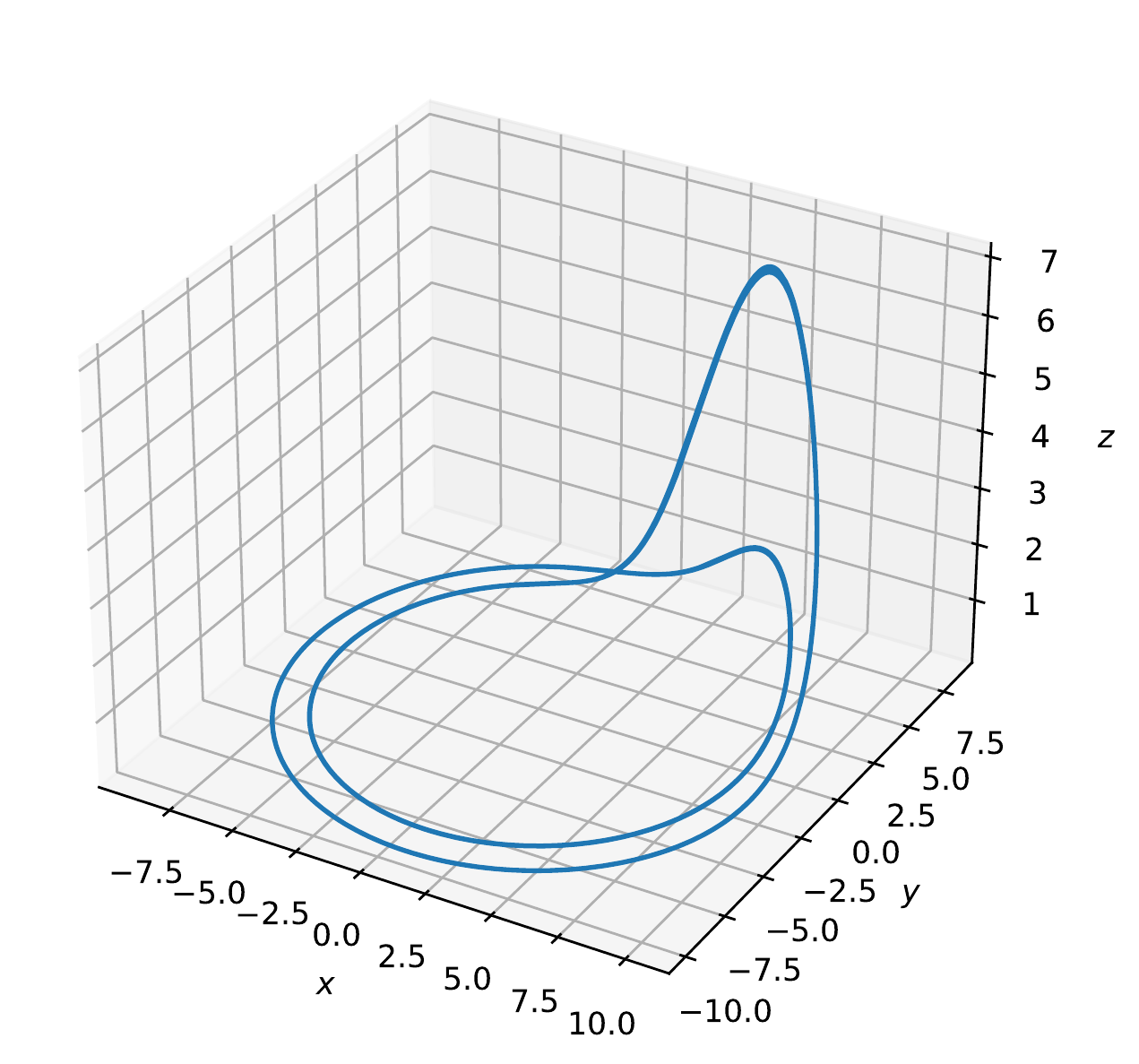}\\
      b)\includegraphics[width=0.9\columnwidth]{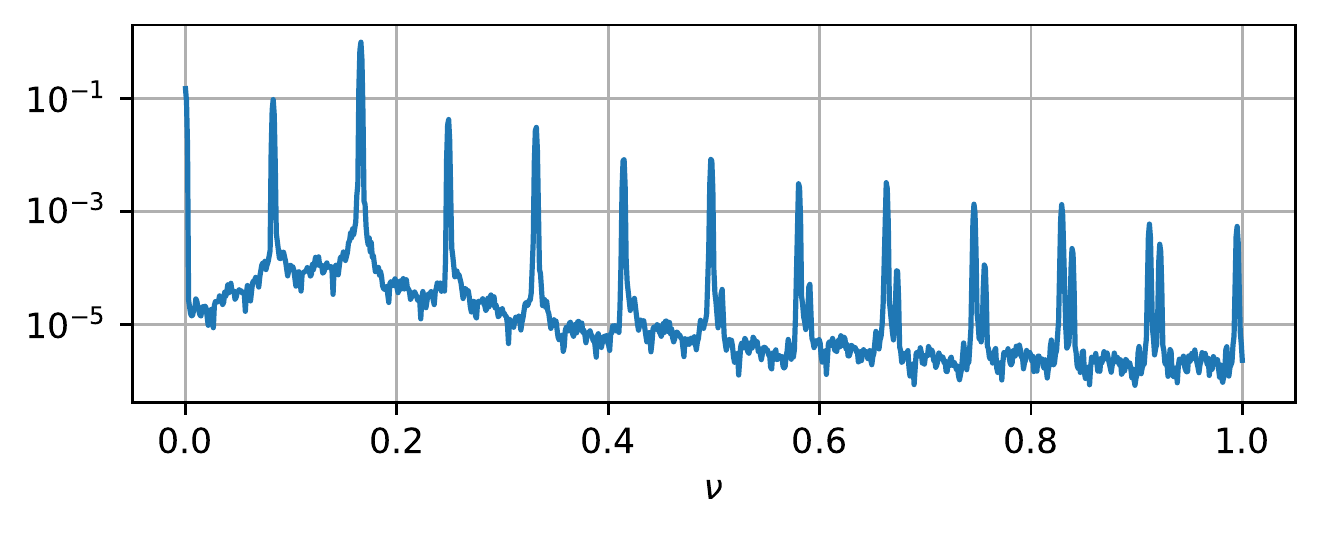}
    \end{minipage} &
    \begin{minipage}{0.5\textwidth}
      c)\includegraphics[width=0.9\columnwidth]{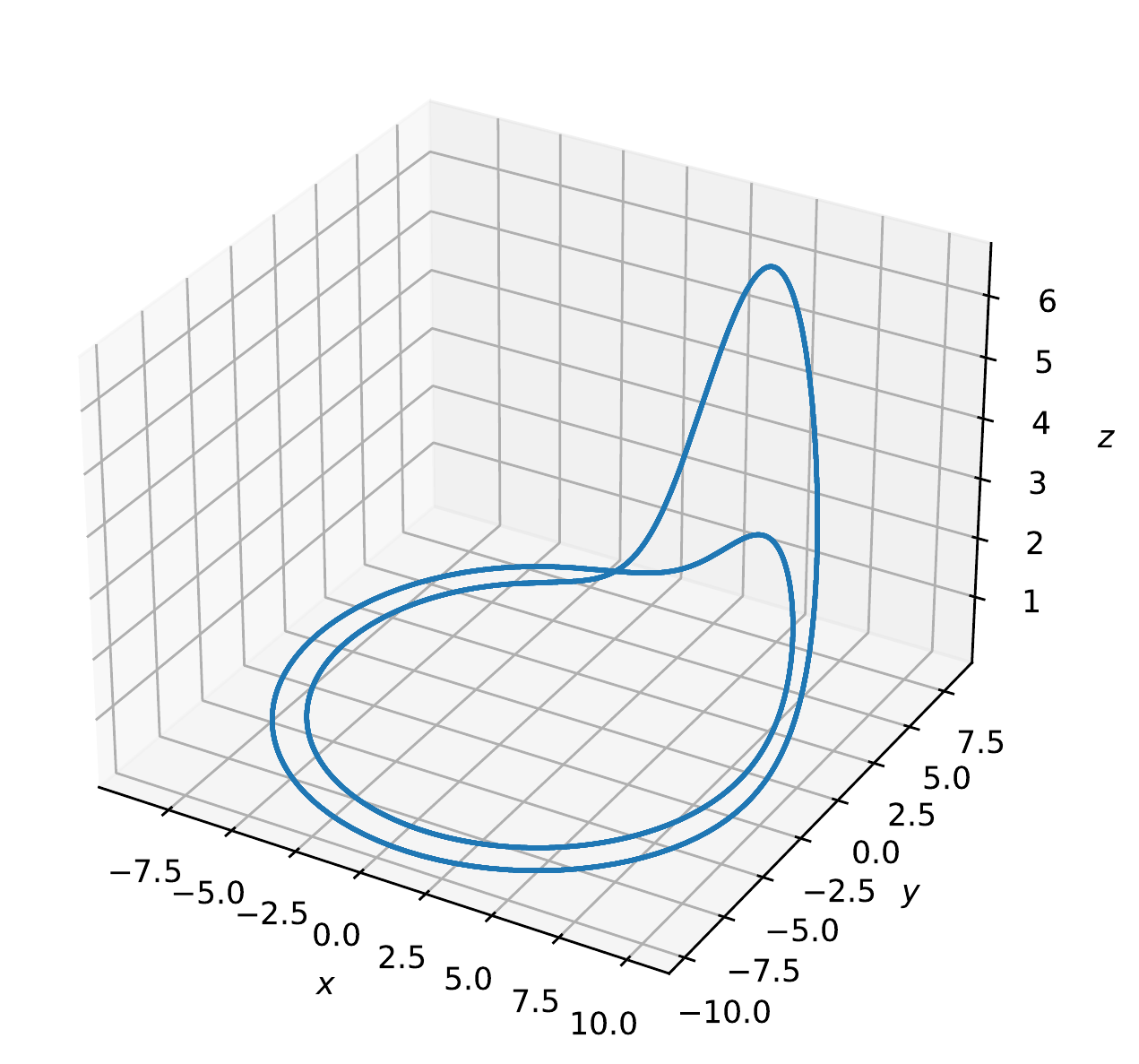}\\
      d)\includegraphics[width=0.9\columnwidth]{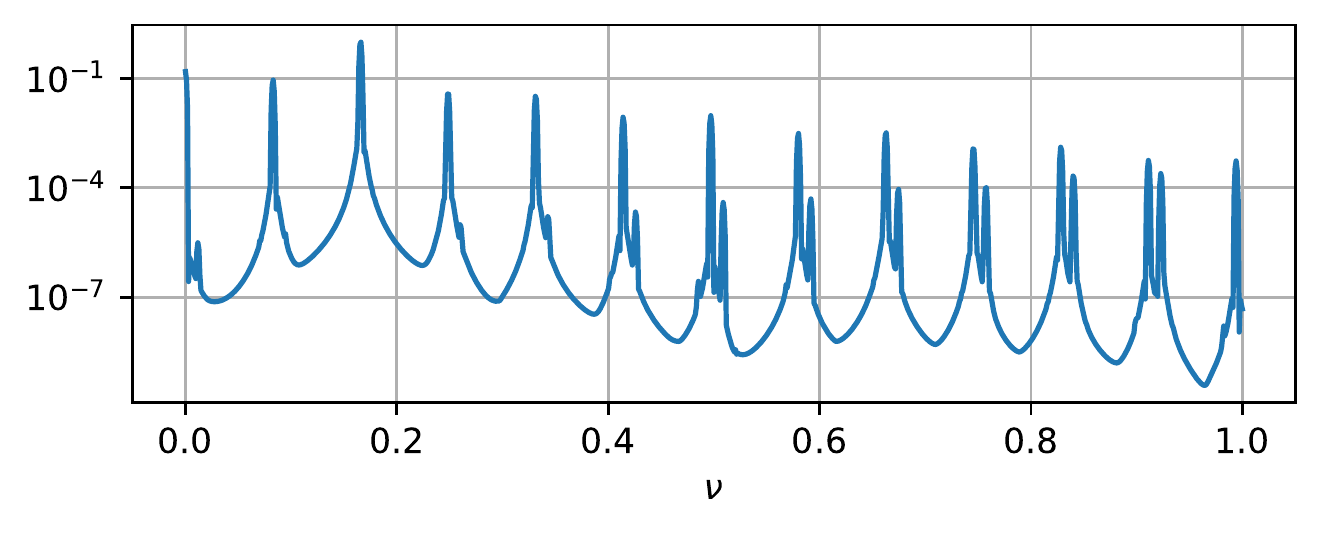}
    \end{minipage}
  \end{tabular}
  \end{center}
  \caption{\label{fig:roessler_attr_four_period}Period 2 oscillations of the \Roes{}
    system at $a=0.1$, $b=0.1$, and $c=6$. Limit cycle (a) and Fourier spectrum
    (b) are computed for ODEs and the corresponding panels (c) and (d) are
    obtained for the network model.}
\end{figure}

Figure~\ref{fig:roessler_attr_four_chaos} demonstrates chaotic \Roes{} attractor
and the corresponding Fourier spectra computed for ODEs~\eqref{eq:roessler_ode}
(left column) and for the trained network model (right column). Numerical
solutions of ODEs here and below are obtained using the forth order Runge-Kutta
method. A very high similarity of the graphs indicates the high quality of
approximation of the network model. Another example of dynamics is in
Fig.~\ref{fig:roessler_attr_four_period}. Parameters here correspond to the
period 2 oscillations. Limit cycles in
Figs.~\ref{fig:roessler_attr_four_period}(a) and (c) looks almost identical. The
spectrum for the network model in Figs.~\ref{fig:roessler_attr_four_period}(d)
also repeats the spectrum in Fig.~\ref{fig:roessler_attr_four_period}(b) in
location and relative heights of harmonics. The difference between these two
spectra is in small fluctuations. Since the regime of the considered system is
periodic the fluctuations are mere artifacts related in particular with the
computation method. The methods of computations are different and so the
fluctuations are.

\begin{figure}
  \centering
  a)\includegraphics[width=0.8\columnwidth]{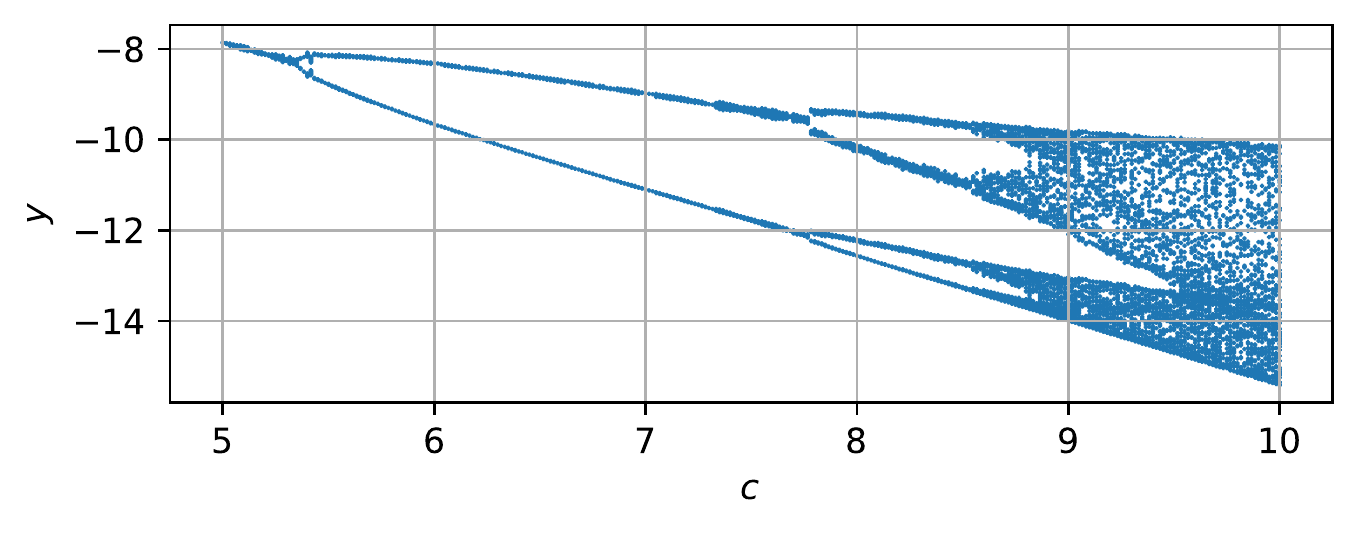}\\
  b)\includegraphics[width=0.8\columnwidth]{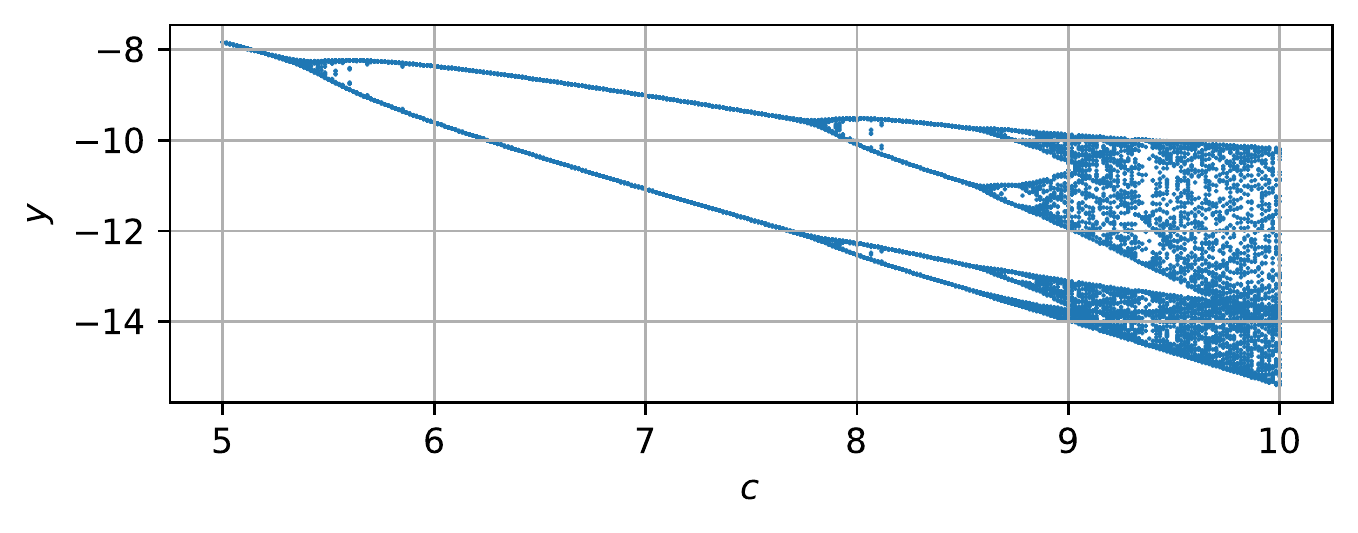}
  \caption{\label{fig:roessler_bifdiag} Bifurcation diagrams for the \Roes{}
    system at $a=0.1$ and $b=0.1$. Panel (a) corresponds to a numerical solution
    of ODEs~\eqref{eq:roessler_ode} and panel (b) is computed for the network
    model. Bifurcation diagrams are obtained as $y$ values at Poincar\'{e}
    sections at $x=0$. The sections are computed for linearly interpolated time
    series.}
\end{figure}

To demonstrate that the trained network model reproduces the dynamics of the
modeled ODEs in a wide range of parameter in Fig.~\ref{fig:roessler_bifdiag} we
show a bifurcation diagram for the \Roes{} system. Parameters $a$ and $b$ are
fixed and $c$ is varying. For each $c$ we compute a trajectory then find its
Poincar\'{e} section at $x=0$. Absent values of variables between the time
discretization points are obtained via linear interpolation. The diagram
obtained for the numerical solution of \eqref{eq:roessler_ode}, see
Fig.~\ref{fig:roessler_bifdiag}(a) is reproduced very well by the network model,
see Fig.~\ref{fig:roessler_bifdiag}(b). Notice however that the bifurcation
points for the network model are little bit shifted to the right. Nevertheless
the overall correspondence is very high.

Let us now compare Lyapunov exponents applying the standard algorithm for
ODEs~\eqref{eq:roessler_ode} and for the corresponding network model. We
demonstrate two cases. For parameters $a=0.15$, $b=0.2$, $c=10$ the Lyapunov
exponents $\lambda_i$ for ODEs are shown in
Eq.~\eqref{eq:roessler_lyap_ode1}. For comparison the exponents
$\tilde\lambda_i$ for the corresponding network model at the same parameters are
shown in Eq.~\eqref{eq:roessler_lyap_ntw1}. The values coincide very
well. Because the considered system is autonomous the value of $\lambda_2$ must
be zero. Actually computed values are indeed very close to zero.
\begin{align}
  \label{eq:roessler_lyap_ode1}
  \lambda_1&=0.0886 & \lambda_2&=-8.66\times 10^{-7} & \lambda_3&=-9.80 \\
  \label{eq:roessler_lyap_ntw1}  
  \tilde\lambda_1&=0.0839 & \tilde\lambda_2&=2.70\times 10^{-5} & \tilde\lambda_3&=-9.64
\end{align}
One more example is considered at $a=0.1$, $b=0.1$, $c=13$ for that the \Roes{}
systems also has a chaotic attractor. From Eqs.~\eqref{eq:roessler_lyap_ode2}
and \eqref{eq:roessler_lyap_ntw2} we again observe that the exponents for the
network model $\tilde\lambda_i$ are close to those obtained for the numerical
solution of ODEs $\lambda_i$.
\begin{align}
  \label{eq:roessler_lyap_ode2}
  \lambda_1&=0.0116 & \lambda_2&=1.87\times 10^{-5} & \lambda_3&=-12.8 \\
  \label{eq:roessler_lyap_ntw2}  
  \tilde\lambda_1&=0.0189 & \tilde\lambda_2&=8.53\times 10^{-5} & \tilde\lambda_3&=-12.8
\end{align}
However we must notice that the correspondence of the Lyapunov exponents for the
\Roes{} system is not so good as for the Lorenz system,
see. Eqs.~\eqref{eq:lorenz_lyap_ode1}-~\eqref{eq:lorenz_lyap_ntw2}. We address
it to the parameter mismatch observed in the bifurcation diagrams.

\subsection{Hindmarch–Rose neuron}

Now we consider the Hindmarsh–Rose model of neuronal
activity~\cite{HindRose84,Wang93}:
\begin{equation}
  \label{eq:hindrose_ode}
  \begin{gathered}
    \dot x = y-a x^3+b x^2-z+I, \\
    \dot y = c-d x^2-y, \\
    \dot z = r(s(x-\alpha)-z)
  \end{gathered}
\end{equation}
Totally this system has eight parameters. However often the system is considered
when six of them have standard values: $a = 1.0$, $b = 3.0$, $c = 1.0$,
$d = 5.0$, $s = 4.0$, $\alpha = -1.6$. Parameters $I$ and $r$ are varied.

The Hindmarch–Rose model~\eqref{eq:hindrose_ode} is a simplified model for
biological neurons presenting bursting oscillations. In this regime, bursts of
fast spikes are followed by quiescent periods. Typical values of parameters
where the bursts are observed are $I=2.7$ and $r=0.003$. Thus we select the
normalization to be close to these values:
\begin{equation}
  \label{eq:hindrose_scl}
  \begin{gathered}
    \mu_u = (0, -5, 2.5), \; s_u = (0.8, 2.5, 0.5), \\
    \mu_p = (0.012, 2.7), \; s_p = (0.024, 0.3), \\
    \mu_g = (-2.9, 2.8, 0.047), \; s_g = (4.2, 5.2, 0.13)
  \end{gathered}
\end{equation}

\begin{figure}
  \centering
  \includegraphics[width=0.96\columnwidth]{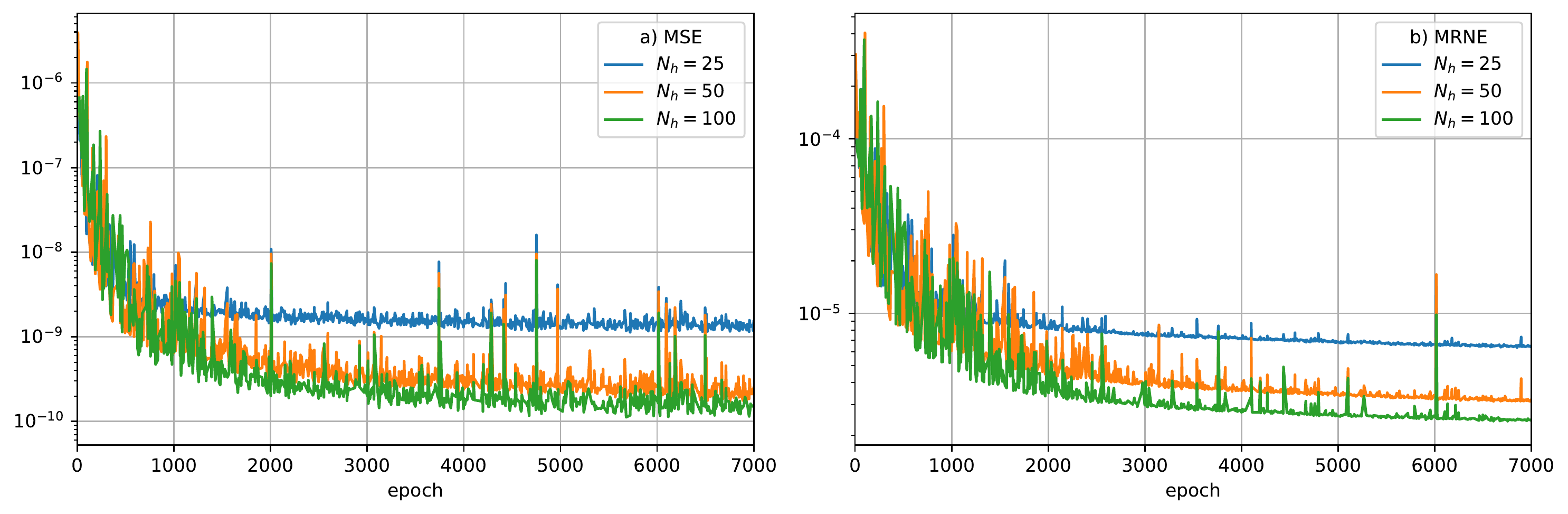}
  \caption{\label{fig:hindrose_learn}Learning curves for the Hindmarch-Rose network
    model corresponding to ODEs~\eqref{eq:hindrose_ode}.}
\end{figure}

Learning curves for the network model of the system~\eqref{eq:hindrose_ode} are
shown in Fig.~\ref{fig:hindrose_learn}. Unlike two previous cases the training
occurs much faster: it takes 7000 epochs for the loss function at $N_h=100$ to
reach values about $10^{-10}$. The model at $N_h=50$ also demonstrates a very
good convergence, and the model at $N_h$ behave much worse. Thus we will
consider a model with $N_h=50$. 

\begin{figure}
  \begin{center}
    a)\includegraphics[width=0.45\textwidth]{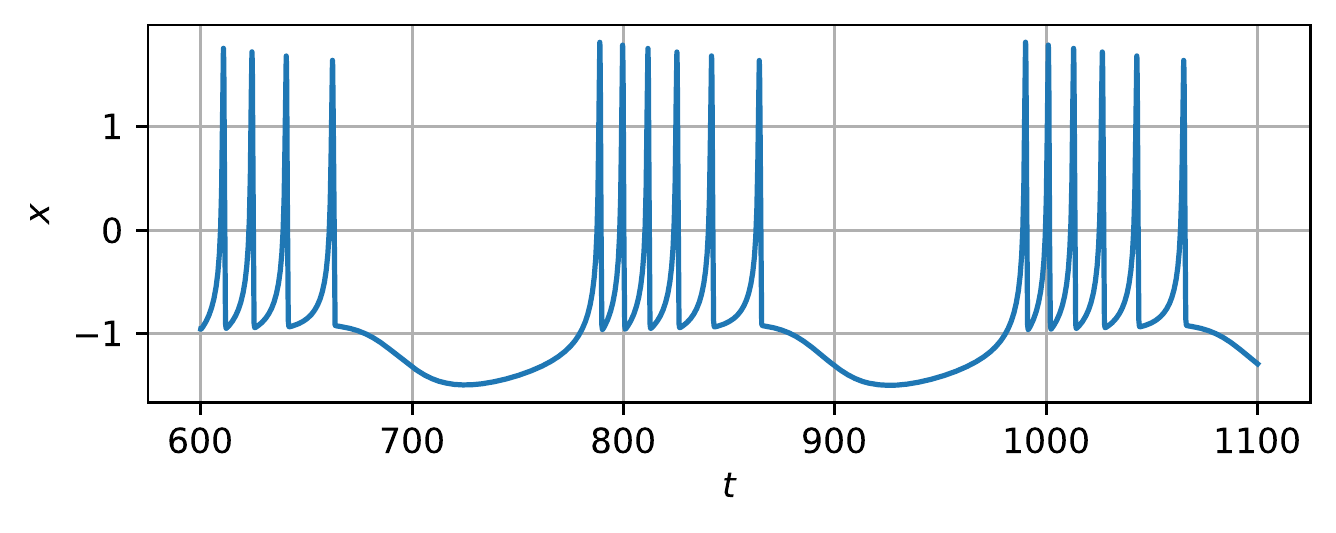}
    b)\includegraphics[width=0.45\textwidth]{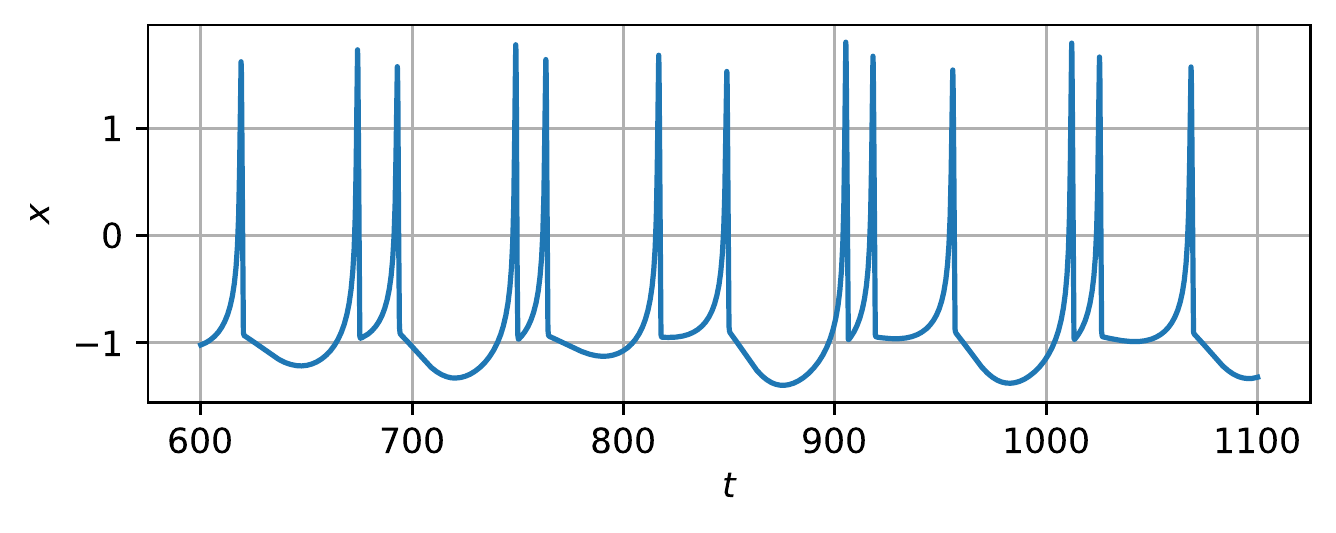}\\
    c)\includegraphics[width=0.45\textwidth]{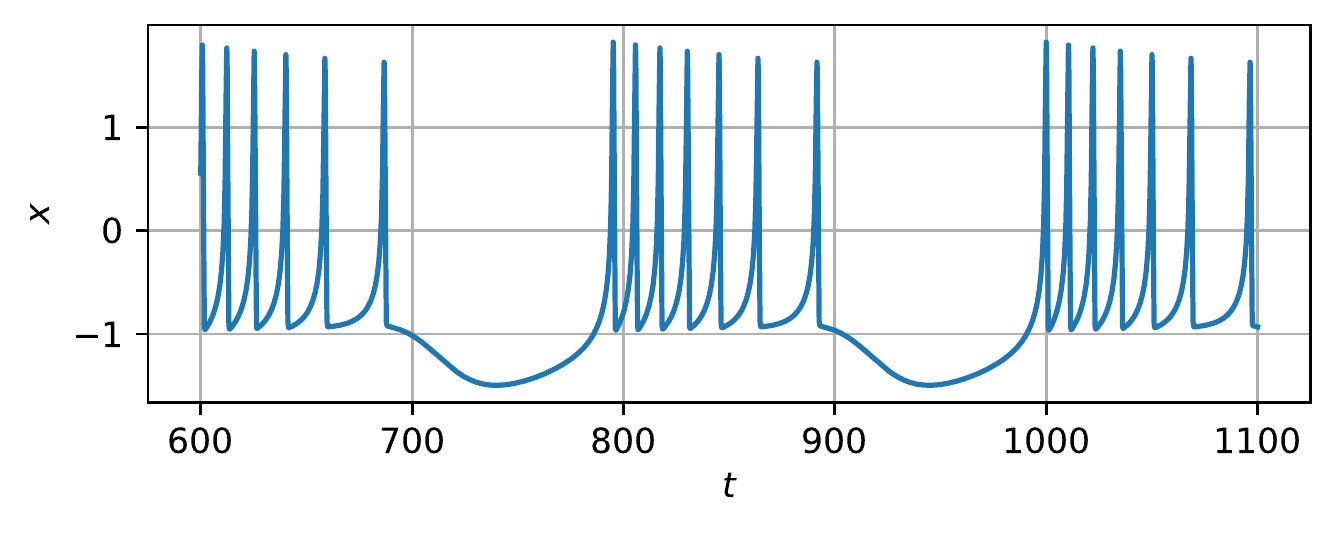}
    d)\includegraphics[width=0.45\textwidth]{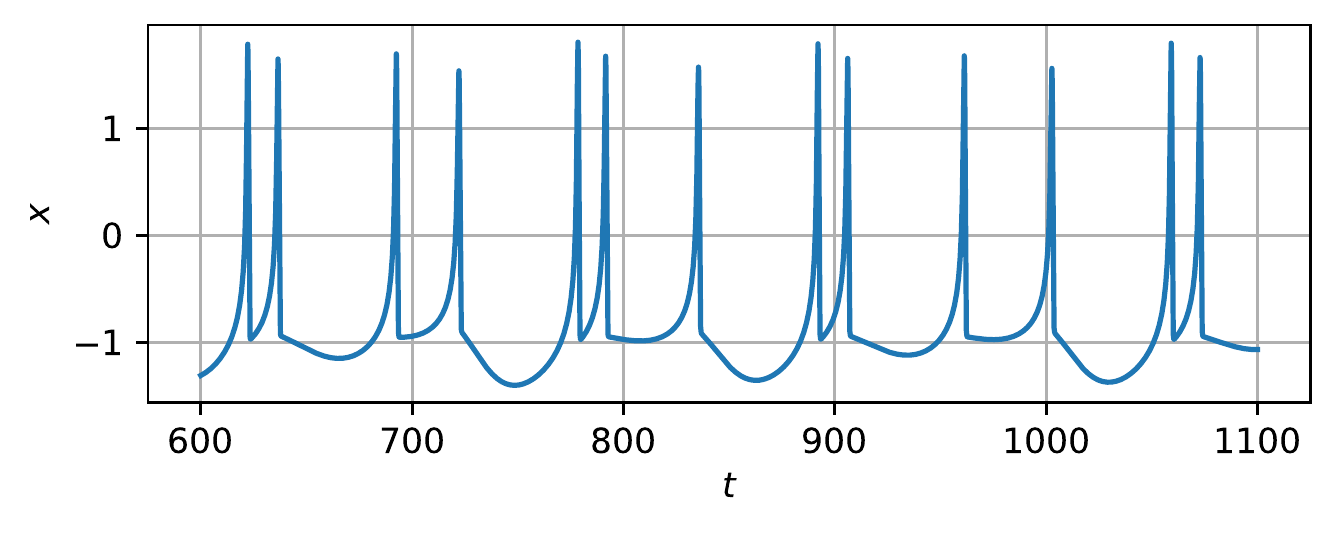}
  \end{center}
  \caption{\label{fig:hindrose_tser}Time series of $x(t)$ obtained as numerical
    solution of ODEs~\eqref{eq:hindrose_ode}, panels (a) and (b), and
    corresponding iterations of the neural network model, panels (c) and
    (d). Parameters are $r=0.003$, $I=2.7$ for panels (a) and (d), and
    $r=0.013$, $I=2.9$ in panels (b) and (d).}
\end{figure}

Figure~\ref{fig:hindrose_tser}(a, b) demonstrates typical solutions of the
Hindmarch-Rose model~\eqref{eq:hindrose_ode}: Panel (a) demonstrates periodical
the bursts of spikes and in the panel (b) we see chaotic spikes. The system
contains fast and slow variables, i.e., is stiff. To solve it numerically the
method LSODA is used~\cite{LSODA}.

Figures~\ref{fig:hindrose_tser}(c, d) shows the corresponding time series
obtained from the neural network model. The behavior of the network model is
very similar, but the close inspection revels that in the panel (c) there are
seven spikes in each burst, while the ``original'' curve contains only six of
them. It means that although the model demonstrates a neural dynamics as well as
in original ODEs, its parameters do not coincide exactly. The chaotic regimes in
the panels (b) and (d) obviously represent the same regime.

\begin{figure}
  \centering
  a)\includegraphics[width=0.8\columnwidth]{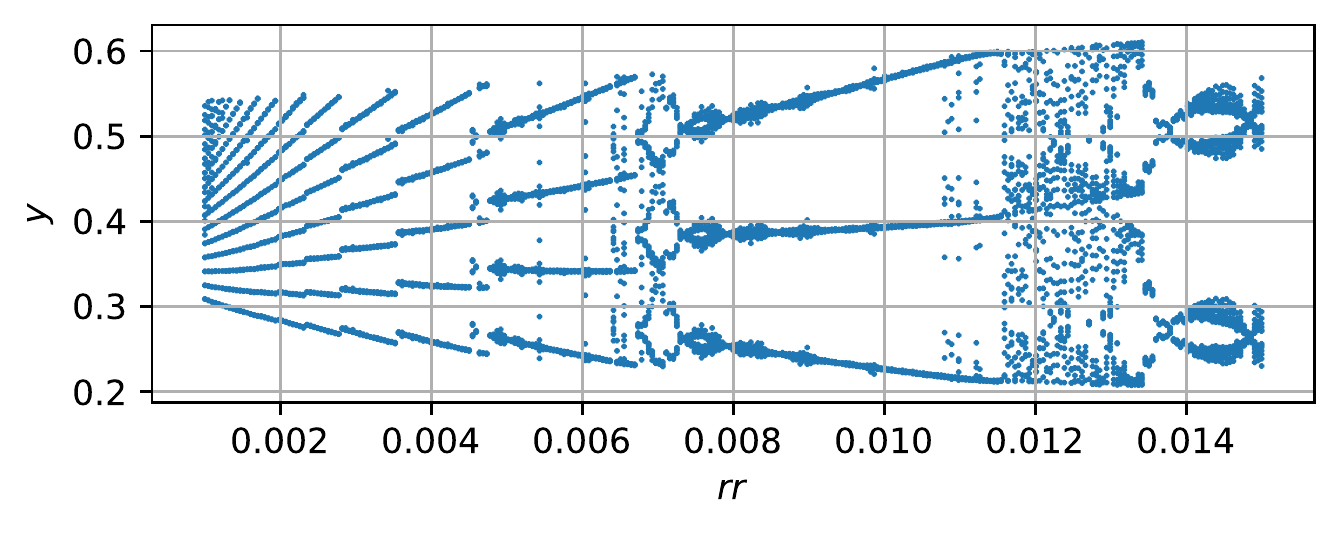}\\
  b)\includegraphics[width=0.8\columnwidth]{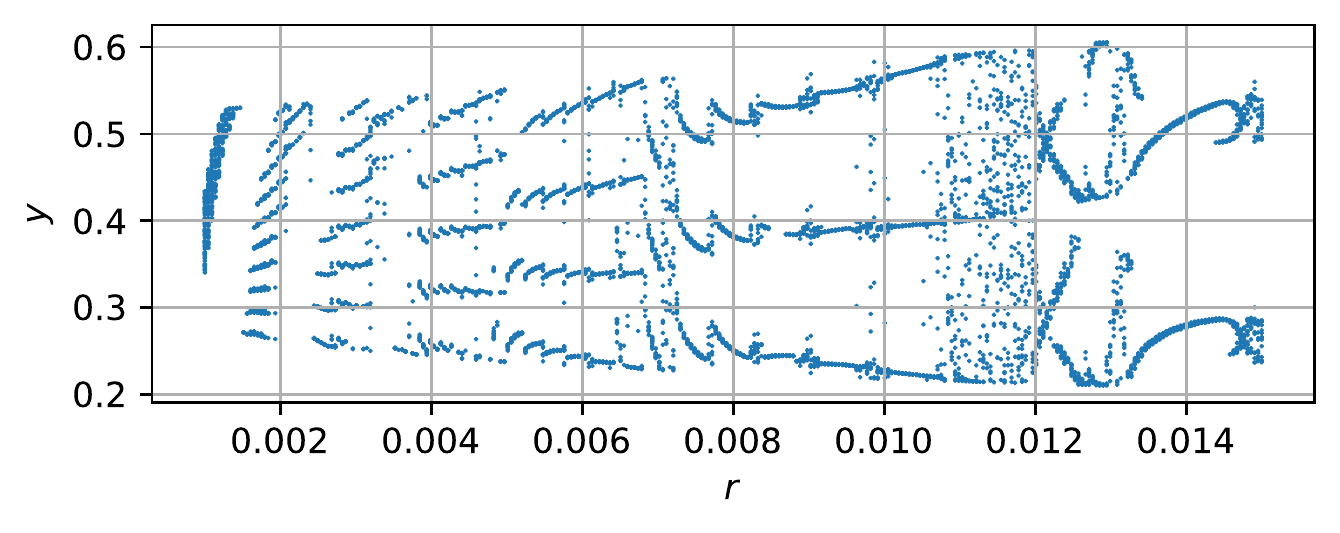}
  \caption{\label{fig:hindrose_bifdiag} Bifurcation diagrams for the \Roes{}
    system at $a=0.1$ and $b=0.1$. Panel (a) corresponds to a numerical solution
    of ODEs~\eqref{eq:roessler_ode} and panel (b) is computed for the network
    model. Bifurcation diagrams are obtained as $y$ values at Poincar\'{e}
    sections at $x=0$. The sections are computed for linearly interpolated time
    series.}
\end{figure}

In Fig.~\ref{fig:hindrose_bifdiag}(a) and (b) bifurcation diagrams provides a
more detailed comparison of the neural network model with ODEs. In both panels
the diagrams are computed for Poincar\'{e} sections at $x=0$ computed for
linearly interpolated times series. The diagrams have similar global
structure. One can see areas of bursts in their left parts and chaotic areas to
the right. However the detailed arrangement is different. The diagram for the
neural network model looks less regular along parameter axis. Often changes of
the regimes are observed. 

Finally we compare Lyapunov exponents computed for numerical solutions of
ODEs~\eqref{eq:hindrose_ode} and for iterations of the neural network model.
For chaotic spikes at $r=0.013$, $I=2.9$, see Fig.~\ref{fig:hindrose_tser}(b),
the Lyapunov exponents $\lambda_i$ are given by
Eq.~\eqref{eq:hindrose_lyap_ode1}. The corresponding exponents for the network
model $\tilde\lambda_i$ are gathered in Eq.~\eqref{eq:hindrose_lyap_ntw1}. The
exponents are pairwise close but do not coincide. 
\begin{align}
  \label{eq:hindrose_lyap_ode1}
  \lambda_1&=8.39\times 10^{-3} & \lambda_2&=-1.32\times 10^{-5} & \lambda_3&=-9.55 \\
  \label{eq:hindrose_lyap_ntw1}  
  \tilde\lambda_1&=7.95\times 10^{-3} & \tilde\lambda_2&=9.77\times 10^{-6} & \tilde\lambda_3&=-9.63
\end{align}
Similar situations is obtained for another parameter values $r=0.012$, $I=2.7$:
the exponents $\lambda_i$ and $\tilde\lambda_i$ are similar, but the difference
is notable.
 \begin{align}
  \label{eq:hindrose_lyap_ode2}
  \lambda_1&=5.41\times 10^{-3} & \lambda_2&=7.59\times 10^{-6} & \lambda_3&=-10.2 \\
  \label{eq:hindrose_lyap_ntw2}  
  \tilde\lambda_1&=3.20\times 10^{-3} & \tilde\lambda_2&=3.04\times 10^{-5} & \tilde\lambda_3&=-10.3
\end{align}
We address this not very good coincides of the Lyapunov exponents to the very
weak chaos. The first exponents are very small by magnitude so that the
numerical routine converges poorly and is strongly affected by numerical errors.
The parameter mismatch observed above for the \Roes{} system also includes in
the not very good coincides of the Lyapunov exponents.

Thus, we observe that the discussed neural network model for the Hindmarch-Rose
system provides good qualitative approximation of this system, however the
quantitative correspondence is not high.

\section{Conclusion}

We discussed the universal neural network, a perceptron with one hidden level,
that can be trained to model behavior of various dynamical systems given by
ODEs. Mathematically the universal neural network model is a discrete time
system, see~\eqref{eq:ode_netw_sol}. We aware of contemporary success in using
of so called deep networks. Our network on contrary is not deep. We have
preferred it because there is a rigorous mathematical evidence, the Universal
Approximation Theorem, that the network with such architecture is able to
approximate various dependencies. Another reason to apply a classical perceptron
is its simple structure. We believe that it will help to trigger new theoretical
studies of dynamical systems. From the practical point of view this simple
network can be effectively simulated using so called AI accelerators, a hardware
dedicated to deal with artificial neural networks. The approach developed in
this paper can be considered as an alternative numerical method of modeling
dynamical system that is able to utilize contemporary parallel hard- and
software.

The universal network model was trained to reproduce the dynamics of the three
systems: Lorenz and the \Roes{} systems and Hindmarch-Rose model. It was very
successful for the Lorenz system. This is confirmed by visual inspection of
attractors, and by coincidence of Fourier spectra and Lyapunov exponents. For
the \Roes{} system the correspondence is also high. However a certain mismatch
of the bifurcation points is observed on bifurcation diagrams computed for the
numerical solution of \Roes{} ODEs and for the network model.

For the Hindmarch-Rose system good qualitative correspondence is achieved
however quantitative characteristics are sometimes differ. This system allows to
reveal the limitations of the suggested approach. Hindmarch-Rose system is stiff
and also its regimes changes fast withing a narrow range parameters. Probably
for such cases like this system a more subtle approach is required.

\bigskip

Work of PVK on theoretical formulation and numerical computations and work of NVS
on results analysis was supported by grant of Russian Science Foundation No
20-71-10048.

\bibliographystyle{unsrt} 
\bibliography{nndyn}

\end{document}